\documentclass[aps,pra,reprint,superscriptaddress,showpacs,floatfix,nofootinbib]{revtex4-1}
\usepackage{hyperref}

\usepackage{graphicx,ulem}
\usepackage{amsmath,amstext,amssymb}
\usepackage{dcolumn}
\usepackage{bm}

\newcommand{\nk}{n_{\rm Kerr}}
\newcommand{\nkk}{n_{\rm Kerr,11}^I}
\newcommand{\ns}{n_{\rm SHG}^I}
\newcommand{\gw}{~{\rm GW/cm^2}}

\newcommand{\lmax}{z_{\rm opt}}

\newcommand{\Ef}{{\mathcal E}}
\newcommand{\neffs}{N_{\rm eff}}%
\newcommand{\nshgb}{N_{\rm SHG}}%
\newcommand{\nkerrb}{N_{\rm Kerr}}%

\newcommand{\deff}{d_{\rm eff}}

\newcommand{\kp}{k^{(1)}}
\newcommand{\kpp}{k^{(2)}}
\newcommand{\tin}{T_{\rm in}}
\newcommand{\Iin}{I_{\rm in}}
\newcommand{\imm}{~{\rm mm^{-1}}}
\newcommand{\topt}{\Delta t_{\rm opt}}

\newcommand{\FT}{{\mathcal F}}

\newcommand{\ie}{{\textit{i.e.}, }}

\newcommand{\mic}{~\mu{\rm m}}
\newcommand{\ev}{{\mathbf e}}

\newcommand{\E}{\mathbf{E}}
\newcommand{\D}{\mathbf{D}}
\newcommand{\Pol}{\mathbf{P}}
\newcommand{\evac}{\varepsilon_0}
\newcommand{\uv}{\mathbf{u}}

\begin{document}


\title{On type I cascaded quadratic soliton compression in lithium
  niobate: Compressing femtosecond pulses from high-power fiber
  lasers}

\author{Morten Bache}\email{moba@fotonik.dtu.dk}
\affiliation{DTU Fotonik, Department of Photonics Engineering,
  Technical University
of Denmark, DK-2800 Kgs. Lyngby, Denmark}%
\author{Frank W. Wise}
\affiliation{Department of Applied and Engineering Physics, Cornell University, Ithaca, New York 14853}

\date{\today}

\begin{abstract}
  The output pulses of a commercial high-power femtosecond fiber laser
  or amplifier are typically around 300-500 fs with a wavelength
  around 1030 nm and 10s of $\mu$J pulse energy. Here we present a
  numerical study of cascaded quadratic soliton compression of such
  pulses in LiNbO$_3$ using a type I phase matching configuration. We
  find that because of competing cubic material nonlinearities
  compression can only occur in the nonstationary regime, where
  group-velocity mismatch induced Raman-like nonlocal effects prevent
  compression to below 100 fs. However, the strong group velocity
  dispersion implies that the pulses can achieve moderate compression
  to sub-130 fs duration in available crystal lengths. Most of the
  pulse energy is conserved because the compression is moderate. The
  effects of diffraction and spatial walk-off is addressed, and in
  particular the latter could become an issue when compressing in such
  long crystals (around 10 cm long). We finally show that the second
  harmonic contains a short pulse locked to the pump and a long
  multi-ps red-shifted detrimental component. The latter is caused by
  the nonlocal effects in the nonstationary regime, but because it is
  strongly red-shifted to a position that can be predicted, we show
  that it can be removed using a bandpass filter, leaving a sub-100 fs
  visible component at $\lambda=515$ nm with excellent pulse quality.
\end{abstract}

\maketitle 

\section{\label{sec:level1}Introduction}

Pulsed fiber laser systems are currently undergoing a rapid
development, and by employing the chirped pulse amplification (CPA)
technique high-energy femtosecond pulses can be generated with
$\mu$J--sub-mJ pulse energies \cite{fermann:2009}. Combined with the
fact that the fiber laser technology offers a rugged, cheap and
compact platform, ultrafast fiber CPA (fCPA) systems could compete
with solid-state amplifier systems. However, the gain bandwidth of the
Yb-doped fibers typically used for lasing in the $1.0\mic$ region is
considerably lower than competing solid-state materials (such as
Ti:Sapphire crystals). Thus, due to the build up of an excessive
nonlinear phase shift Yb-based fCPA lasers are often limited to a
pulse duration that typically is sub-ps at best (around $500-700$ fs)
for $\sim 100~\mu$J pulses \cite{limpert:2006} while shorter pulses can
be reached ($\sim 250$ fs) for $\sim 30~\mu$J pulses
\cite{Kuznetsova:2007}.

Efficient external compression methods are therefore needed. A
prototypical compressor consists of a piece of nonlinear material,
where a broadening of the pulse bandwidth occurs by self-phase
modulation (SPM), followed by a dispersive element (gratings or
chirped mirrors) that provides temporal compression.  
With this method (using a short
piece of fiber as nonlinear material) 27 fs sub-$\mu$J pulses were
generated from 270 fs 0.8 $\mu$J pulses from an fCPA system
\cite{eidam:2008}.  
Alternative methods consist of using
long (0.5 m or more) gas cells or filaments
\cite{nisoli:1996,*hauri:2004} as nonlinear material, and this works with pulse
energies from 50 $\mu$J to around 1 mJ (limited in part by
self-focusing effects) or possibly even higher energies \cite{Chen:2008}. 

Using soliton compression both the SPM-induced pulse broadening and
dispersion-induced compression occur in the same material
\cite{mollenauer:1980}. However, as self-focusing solitons require
anomalous dispersion this can only be achieved in the near-IR through strong
waveguide dispersion. This means using specially designed fibers, such
as micro-structured fibers. Fibers have a very limited maximum pulse
energy of a few nJ, albeit large mode-area micro-structured solid-core
and hollow-core fiber compressors can support up to 1 $\mu$J
\cite{Ouzounov:2005,*Laegsgaard:2009}.

\begin{figure}[b]
\includegraphics[width=6cm]{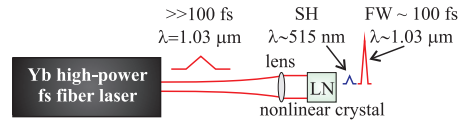}
\caption{\label{fig:setup} (Color online) The cascaded
  quadratic soliton compressor studied here: the Yb fiber laser produces
  energetic longer pulses ($\gg 100$ fs) that are launched
  collimated in a quadratic nonlinear lithium niobate crystal, where
  the phase-mismatched type I SHG process compresses the input pulse.}
\end{figure}

Unfortunately the pulse energy from fCPA systems lies
exactly in the gap between these methods. We will here study a
compression method that can compensate for this. It is a soliton
compressor based on cascaded quadratic nonlinearities
\cite{liu:1999,ashihara:2002,wise:2009}, see Fig.~\ref{fig:setup}. This has
several advantages: As it relies on a self-defocusing nonlinearity,
there are no problems with self-focusing effects, and multi-mJ
pulse-energies can be compressed. Moreover, solitons require normal
instead of anomalous dispersion, implying that solitons can be
generated in the visible and near-IR.  Finally, it is extremely simple
as it relies on just a small piece of quadratic nonlinear crystal,
preceded only by a lens or a beam expander \cite{moses:2007}.

The basis for the cascaded quadratic soliton compressor (CQSC) is
phase-mismatched second-harmonic generation (SHG). The cascaded
energy transfer from the pump (fundamental wave, FW) to the second
harmonic (SH) and back imposes a strong SPM-like nonlinear phase shift
on the FW, whose sign can be made self-defocusing
\cite{desalvo:1992,stegeman:1996}. Thereby the FW pulse can be
compressed with normal dispersion \cite{liu:1999}, and soliton
compression becomes possible in the visible and near-IR
\cite{ashihara:2002}.

In this paper we investigate the CQSC in a type I lithium niobate
(LiNbO$_3$, LN) crystal, where
the goal is to perform moderate compression of longer fs pulses from
fCPA systems at the Yb gain wavelength of 1030 nm. We show that in
order to overcome the detrimental cubic nonlinearities the phase
mismatch has to be chosen so low so that the compression occurs in the
so-called nonstationary regime.  This regime is dominated by group-velocity
mismatch (GVM) effects, and
exactly the large GVM is a well-known drawback of using LN in the
near-IR for SHG. However, when only moderate compression is desired,
the soliton order can be kept low, and we show through numerical
simulations that reasonable pulse quality can be achieved and that up
to 80\% of the pulse energy is retained in the central spike. The
compression limit is found to be around 120 fs FWHM, which is a limit
set by the GVM effects. The compression occurs in a crystal of
reasonable length, 10 cm. This is possible only because LN has a very
large 2. order dispersion.  Finally, we show that bandpass filtering
of the SH actually can lead to a very clean sub-100 fs visible pulse
with around 0.1\% conversion efficiency.

In this paper we first discuss the general
compression properties of LN in a cascaded type I SHG interaction
setup in Sec.~\ref{sec:Compr-prop-lith}, and then show some numerical
simulations in Sec.~\ref{sec:Numer-simul} of pulses coming from two
different commercially available fCPA systems. We conclude in
Sec.~\ref{sec:Conclusion}. The properties of LN are discussed in
App.~\ref{sec:Crystal-parameters}, and App.~\ref{sec:Anis-Kerr-nonl}
discusses the anisotropic Kerr nonlinear response of
LN. Appendix~\ref{sec:Conversion-relations}
and~\ref{sec:Wavel-scal-nonl} discuss the conversion relations between
Gaussian and SI units for cubic nonlinear coefficients and Miller's
rule, respectively.

\section{Type I compression properties of lithium niobate 
  crystals}
\label{sec:Compr-prop-lith}

With the CQSC high-energy few-cycle compressed pulses can be
generated, as was experimentally observed at 1250 nm
\cite{moses:2006}.  However, the first studies performed at 800 nm
were plagued by GVM effects, that prevented reaching the few-cycle
regime \cite{liu:1999,moses:2006,ashihara:2002}.  These studies used a
$\beta$-barium--borate (BBO) crystal in a type I SHG $oo\rightarrow e$
configuration, where the FW (ordinary polarization) is orthogonal to
the SH (extraordinary polarization) and where birefringent phase
matching is possible by angle-tuning the crystal. BBO is in many
respects an ideal nonlinear crystal: it has low dispersion, a very
large transparency window, and a reasonably strong quadratic
nonlinearity relative to the detrimental cubic one.  As we have shown
in previous theoretical and numerical studies, BBO provides an
excellent compression of longer pulses to ultra-short duration at the
Yb gain wavelengths \cite{bache:2007a,bache:2007,bache:2008}.  The
problem with BBO is that good quality waveguides are not supported and
that it is very difficult to grow long crystals. Especially the latter
is important if only moderate compression of longer pulses is desired.
In moderate soliton compression most of the pulse energy is conserved
in the compressed pulse, and the pulse has a reduced pedestal. The
problem is that compression will only occur after a long propagation
length.

We therefore turn here to LN, which is a widely used quadratic
nonlinear crystal for IR frequency conversion. LN is attractive due to
extremely large effective quadratic nonlinearities (up to 10 times
larger than BBO), that can be accessed through a quasi-phase matched
(QPM) type 0 SHG phase matching configuration where FW and SH have
identical polarization.  However, here we study LN in a type I
configuration as BBO. The effective quadratic nonlinearity is more
than twice as large as in BBO.

LN is usually not considered very suitable for SHG of short pulses in
the near-IR because the SH becomes very
dispersive; thus, the FW and SH group velocities are very different
resulting in large GVM. This is also why LN has not been used
in the near-IR as nonlinear medium for the CQSC, for which GVM is a
very detrimental effect.  Another disadvantage for the CQSC is that
the Kerr nonlinear response is several times larger than BBO, which
counteracts the advantage of the large quadratic nonlinearity of LN.
Therefore the CQSC experiments done so far using LN were done in the
telecommunication band and exploited QPM in a type 0 configuration
\cite{ashihara:2004,*zeng:2006}, where effective quadratic nonlinearity
is around three times larger than what can be achieved in a type I
configuration. However, we now show that type I LN offers a
quite decent compression performance without having to custom design a
QPM grating.

\subsection{Solitons with cascaded quadratic nonlinearities}
\label{sec:Type-I-cascaded}

In cascaded quadratic interaction the FW effectively experiences a
Kerr-like nonlinear refractive index. This is in addition to the cubic
(Kerr) nonlinearities that are always present in all media. We can
write the total refractive index of the FW [see
Eq.~(\ref{eq:n2-def})]
\begin{eqnarray}
  n=n_1+\tfrac{1}{2}|\Ef_1|^2 n_{\rm
  cubic}=n_1 + I_1 n_{\rm cubic}^{I}
\end{eqnarray}
where $n_1$ is the FW linear refractive index, $\Ef_1$ is the FW
electric field, and $I_1$ the FW intensity. It is typical to report
the nonlinear refractive index relative to the electric field, $n_{\rm
  cubic}$, or to the intensity, $n_{\rm cubic}^{I}$. We have here for
simplicity neglected cross-phase modulation (XPM) contributions since
they are small in cascaded SHG. As mentioned we have contributions
from both cascaded quadratic and cubic Kerr nonlinearities
\begin{eqnarray}\label{eq:cubic}
  n_{\rm cubic}^{I}=n_{\rm SHG}^I+n_{\rm Kerr,11}^I
\end{eqnarray}
where $n_{\rm Kerr,11}^I$ is the SPM Kerr nonlinear refractive index
of the FW (see App.~\ref{sec:Anis-Kerr-nonl} for details on the
notation etc.). The contribution from the cascaded quadratic
nonlinearities can in the large phase mismatch limit ($\Delta k L\gg
1$, where $L$ is the crystal length) be approximated
as \cite{desalvo:1992}
\begin{eqnarray}\label{eq:n2-SHG}
  n_{\rm SHG}^I\simeq -\frac{4 \pi d_{\rm
  eff}^2}{c\varepsilon_0\lambda_1n_1^2n_2 \Delta k}
\end{eqnarray}
where 
$d_{\rm eff}$ is the effective $\chi^{(2)}$ nonlinearity. For $\Delta
k=k_2-2k_1>0$ the cascaded contribution is negative, i.e. self-defocusing. Here
$k_j=2\pi/\lambda_j$ is the wavenumber. 

The effective quadratic nonlinearity of the type I $oo\rightarrow e$
interaction for the $3m$ crystal class (LN, BBO) is
\begin{eqnarray}\label{eq:deff}
  d_{\rm eff}=d_{31}\sin\theta-d_{22}\cos\theta\sin3\phi
\end{eqnarray}
where the angles are defined in Fig.~\ref{fig:crystal} in
App.~\ref{sec:Anis-Kerr-nonl}. Choosing $\phi=-\pi/2$ gives maximum
nonlinearity (see App.~\ref{sec:Crystal-parameters}).

In cascaded quadratic soliton compression the aim is to get $\ns<0$
and $|\ns|>\nkk$ as to achieve a total self-defocusing cubic
nonlinearity. The soliton interaction can then be described by an
effective soliton order \cite{bache:2007}
\begin{eqnarray}\label{eq:Neff}
  \neffs^2&=&\nshgb^2-\nkerrb^2\\
&=&L_{D,1}k_1\Iin(|\ns|-\nkk)\nonumber
\end{eqnarray}
where $\nshgb=L_{D,1}k_1\Iin|\ns|$ is the soliton order of the
self-defocusing cascaded quadratic nonlinearity, and
$\nkerrb=L_{D,1}k_1\Iin\nkk$ is the soliton order of the material Kerr
self-focusing cubic nonlinearity. The FW dispersion length is $L_{\rm
  D,1}=\tin^2/|\kpp_1|$, where $\kpp_1$ is the FW group-velocity
dispersion (GVD). We generally use the following notation for the
dispersion parameters $k_j^{(m)}=\partial^m k_j/\partial
\omega^m |_{\omega=\omega_j}$.
 
\begin{figure}[tb]
\includegraphics[width=8.5cm]{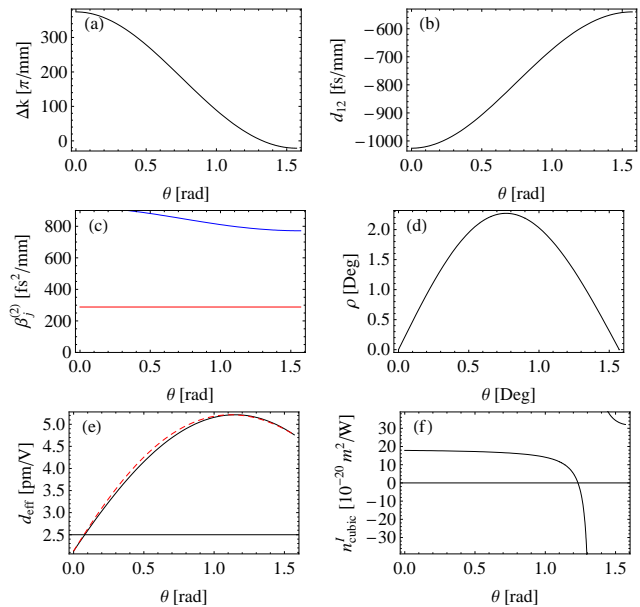}
\caption{\label{fig:1060nm} (Color online) Properties at
  $\lambda_1=1.03\mic$ when angle-tuning the LN crystal: (a) Phase
  mismatch, (b) GVM parameter, (c) GVD of FW (red) and SH (blue), and
  (d) the spatial walk-off angle $\rho$.  The effective quadratic
  nonlinearity neglecting (black) and including (dashed red) spatial
  walk-off are shown in (e) and (f) is the total cubic Kerr
  nonlinearity~(\ref{eq:cubic}) from cascaded quadratic nonlinearities
  and Kerr SPM (using $\nk^I=18\times 10^{-20}~\rm m^2/W$, see
  App.~\ref{sec:Anis-Kerr-nonl}).}
\end{figure}

\subsection{Linear and nonlinear response of LN at $1.03\mic$}
\label{sec:Compression-LN-at}

Selecting $\lambda_1=1.03\mic$, the operating wavelength of most
Yb-based fiber laser amplifiers, the properties of LN are summarized
in Fig.~\ref{fig:1060nm}: the phase mismatch (a) becomes small at
$\theta\simeq 1.3$ radians ($70-75^\circ$). As shown in (e) in this
range $\deff\simeq 5.2$ pm/V, and the total nonlinear refractive index
(f), as expressed by Eq.~(\ref{eq:cubic}), can become negative,
implying that the cascaded nonlinearity is stronger than the Kerr
nonlinearity. This happens for $\Delta k<62\imm$ (or
$\theta>70.4^{\circ}$). At $\theta=75.8^\circ$ phase matching
is achieved, after which $\ns>0$ and thus self-focusing. 

GVM is very large, see Fig.~\ref{fig:1060nm}(b), which as we will see
later sets a strong limitation to the compression performance. The GVD
is shown in Fig.~\ref{fig:1060nm}(c), and importantly FW GVD (red) is
large and normal (i.e. positive). It will stay normal until
$\lambda_1>1.9~\mu$m, after which it becomes anomalous and
self-defocusing solitons are no longer supported. The SH GVD (blue) is
about 3 times larger than the FW GVD.

Since the type I critical phase matching is employed, the walk-off
angle $\rho=\arctan[\tan (\theta) n_o^2/n_e^2]-\theta$ (valid for a
negative uniaxial crystal) is nonzero, see Fig.~\ref{fig:1060nm}(d).
In Fig.~\ref{fig:1060nm}(e) it is apparent that $\deff$ is largely
unaffected by walk-off. However, walk-off does set a limit to the
effective interaction length between the pump and the SH as we will
discuss later.


\begin{figure}[tb]
\includegraphics[width=8cm]{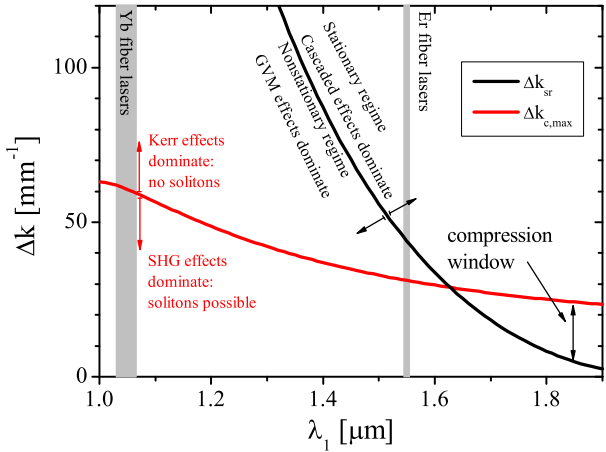}
\caption{\label{fig:LN-cw} (Color online) Compression diagram for 1\%
  MgO:sLN at room temperature and aligned for type I SHG. For various
  pump wavelengths $\lambda_1$ the choice of phase-mismatch parameter
  $\Delta k$ affects the compression. In order to excite solitons the
  phase-mismatch must be kept below the red line ($\Delta k<\Delta
  k_{c,\rm max}$), because otherwise the material cubic nonlinearities
  are too strong ($n_{\rm Kerr,11}^I>|\ns|$). Optimal compression
  occurs when the cascaded nonlinearities dominate over GVM effects
  ($\Delta k>\Delta k_{\rm sr}$, above the black line).  We have also
  indicated the operation wavelengths of Yb and Er doped fiber lasers.
  The red line uses Miller's rule to estimate the nonlinear quadratic
  and cubic susceptibilities at other wavelengths, cf.
  Eqs.~(\ref{eq:Miller-chi2})-(\ref{eq:Miller-chi3}), and uses $n_{\rm
    Kerr,11}^I=20\times 10^{-20}~{\rm m^2/W}$ for $\lambda=0.78\mic$
  (see App.~\ref{sec:Anis-Kerr-nonl} for an extended discussion).}
\end{figure}

\subsection{Compression diagram for type I LN}
\label{sec:Compr-diagr-type}

We now generalize to other wavelengths
and summarize the type I compression performance of LN in
Fig.~\ref{fig:LN-cw} \footnote{The specific crystal chosen in this
  work is 1\% MgO doped stoichiometric LN, as the MgO doping gives a
  much higher material damage threshold. Also 5\% MgO doped congruent
  LN would work well. See
  App.~\ref{sec:Crystal-parameters} for more details about the
  crystal.}. This compression diagram shows the different compression
regimes for the CQSC as the wavelength and the phase mismatch is
varied.

Above the red curve the total nonlinear refractive index is focusing $n_{\rm
  cubic}^I>0$, so solitons are not supported since the FW GVD is normal.
The curve is found by setting $|\ns|=\nkk$ giving \cite{bache:2007}
\begin{eqnarray}
\label{eq:kcmax}
  \Delta k_{c,\rm max}=
k_1\frac{2 d_{\rm
  eff}^2}{c\varepsilon_0n_1^2n_2 \nkk}
\end{eqnarray}
Below the black curve the compression performance is dominated by GVM
effects (nonstationary regime) while above it is dominated by cascaded
effects (stationary regime). The curve is to second order
\footnote{A more accurate transition can
  easily be calculated numerically using the full SH dispersion
  operator \cite{bache:2008}, which we have done in what follows.}
given by \cite{bache:2007a}
\begin{eqnarray}
  \label{eq:ksr}
\Delta k_{\rm sr}=\frac{d_{12}^2}{2\kpp_2}
\end{eqnarray}
where $d_{12}=\kp_1-\kp_2$ is the GVM parameter and $\kpp_2$ is the SH
GVD. The lower this curve is the better because this implies that the
chance of observing solitons in the stationary regime increases. Thus,
the very large GVM parameter $d_{12}$ is detrimental because it pushes
the curve upwards. Instead the huge SH GVD values, see
Fig.~\ref{fig:1060nm} (c), are actually helping to push the curve
downwards. Therefore a large SH GVD can actually be beneficial for
clean soliton compression.

The optimal compression occurs in the so-called ``compression window''
\cite{bache:2007a}, where the soliton compressor works most
efficiently because solitons are supported in the stationary regime.
The diagram shows a compression window for type I LN in the regime
$\lambda_1=1.6-1.9~\mu$m.  Unfortunately in this range there are no
fCPA
systems. 

Fortunately, as we will show also in the nonstationary regime
compression is possible, as long as the effective soliton order is low
enough. This is what we will try to exploit in the regime around $
\lambda_1\sim 1.03-1.06~\mu$m. 

Coming back to $\lambda_1=1.03\mic$ we observe that solitons are
supported for when $0\ll\Delta k<\Delta k_{c,\rm max}=62 \imm$.
However, when getting too close to $\Delta k_{c,\rm max}$ the
intensities required to observe solitons become very large implying
excessive Kerr XPM effects and increased Raman-like GVM effects
\cite{bache:2008}. On the other hand for $\Delta k$ too small the
cascading limit ceases to hold, and also the compressor performance
decreases due to excessive GVM effects \cite{bache:2008}.  In fact, as
a rule of thumb the compression limit in the nonstationary regime (in
which the system will always be for $\Delta k\sim 0$) the compression
limit is roughly given by the pulse duration for which $L_{\rm
  coh}=L_{\rm GVM}$, where $L_{\rm coh}=\pi/|\Delta k|$ is the
coherence length and $L_{\rm GVM}=\Delta t_{\rm soliton}/|d_{12}|$ is
the \textit{dynamic} GVM length of a sech-shaped soliton. With
``dynamic'' we mean that the GVM length changes as the soliton
compresses. Thus, in the nonstationary regime the limit
is \footnote{Note that this expression differs with a factor of
  $\pi/2$ from the limit $T_{R\rm,SHG}=2|d_{12}/\Delta k|$ that we
  suggested in \cite{bache:2008}; this is purely an empirical choice.}
\begin{eqnarray}\label{eq:tlimit}
  \Delta t_{\rm limit}^{\rm FWHM}\sim 2\ln(1+\sqrt{2}) \frac{\pi
    |d_{12}|}{|\Delta k|}
\end{eqnarray}
where the factor in front of the fraction is the conversion factor to
FWHM for a sech-shaped pulse. Obviously as $\Delta k$ approaches the
phase matching point the soliton cannot compress to short durations.
We numerically found the optimal compression point in the $\Delta
k=35-50\imm$ regime, and with the best results for $\Delta k=45\imm$,
for which $\ns=25\times 10^{-20}~\rm m^2/W$.

\subsection{Predicting the compression performance}
\label{sec:Pred-compr-perf}

The next step is to estimate what the compression performance could
look like. Here the scaling laws \footnote{Note that the scaling laws
  presented here are only ball-park figures when used in the
  nonstationary regime as they were found in the stationary regime.}
come into the picture, which can be used to predict the propagation
distance for optimal compression $z_{\rm opt}$, the compression factor
$f_c$ and the pulse quality $Q_c$ \cite{bache:2007}.

As we have pointed recently \cite{bache:2007a}, it is the phase
mismatch and the GVM (zero and first order dispersion) that really
control the compression properties. The only requirement to the second
order dispersion is that FW GVD is normal $\kpp_1>0$ as to support
solitons. Otherwise as we discuss below the FW GVD is basically just
determining the optimum compression length. The SH GVD instead plays a
minor role in the compression properties, cf.  Eq.~(\ref{eq:ksr}). Our
initial idea was to exploit that LN is quite dispersive when pumped at
$ \lambda_1\sim 1.0~\mu$m, so the very large FW GVD makes it possible
to compress the pulse in a short crystal.

So why and when is it interesting to increase GVD as to compress in a
short crystal?  Obviously, the crystals have length limits, which for
LN is around 100 mm. The optimal compression point scales as
\cite{bache:2007} 
\begin{eqnarray}
  \label{eq:zopt-fit}
\frac{\lmax}{z_0}=\frac{0.44}{\neffs}+\frac{2.56}{\neffs^3}-0.002.
\end{eqnarray}
where $z_0=\tfrac{\pi}{2} L_{\rm D,1}$ is the soliton length
\cite{agrawal:1989}. So the point where the pulse compression is
optimal depends on the effective soliton order, the input pulse
duration and the FW GVD. Therefore since quality LN crystals are
maximum 100 mm long, the CQSC works best when the soliton order is
large and the GVD length is short. But when the soliton order is
large, the detrimental effects due to GVM are strongly increased
\cite{moses:2006,bache:2008}, in particular in the nonstationary
regime. Therefore, in the case we study here clean compression can
only be done with low soliton order, and therefore the FW GVD
must be large as to ensure compression in realistic crystal lengths.

A downside to the large GVD is the following: given that some
effective soliton order is required then since $\neffs\propto
\sqrt{\Iin L_{\rm D,1}}\propto \tin\sqrt{\Iin/|\kpp_1|}$ we have that
a large GVD gives a short GVD length, and thus larger intensities are
needed to excite a soliton.  The same problem is found for short input
pulses, say from a Ti:Sapphire amplifier. However, this is only an
issue if operating with intensities close to the damage threshold,
which is not the case here: the intensities are moderate ($\Iin\ll
100\gw$), and instead our issue is to get the solitons to compress
in a crystal that is not too long. 

The compression factor $f_c=\tin/\topt$, where $\topt$ is the pulse
compressed pulse duration at $z_{\rm opt}$, is also affected by the
effective soliton order \cite{bache:2007}
\begin{eqnarray}
  \label{eq:fc-fit}
f_c=4.7(\neffs-0.86)
\end{eqnarray}
The pulse quality can also be predicted, and is defined as the ratio
between the compressed pulse fluence with that of the input pulse. It
scales as \cite{bache:2007}
\begin{eqnarray}
  \label{eq:qc-fit}
Q_c=[0.24 (\neffs-1)^{1.11}+1]^{-1}.
\end{eqnarray}
We can use this to calculate the compressed pulse peak intensity
$I_{\rm opt}=Q_c f_c \Iin$ and energy $E_{\rm opt}=Q_c E_{\rm in}$. An
advantage of using low soliton orders is that $Q_c$ remains high, and
thus the compressed pulse retains most of the initial pulse energy.

\subsection{Compression performance of fCPA systems}
\label{sec:Comm-fCPA-syst}

Let us use these scaling laws to predict the compression performance
of fCPA systems.  High-energy femtosecond pulses from fCPA systems use
both Yb doped and Er doped gain fibers. Since fCPA systems are diode
pumped with a wavelength just below $1.0\mic$ the quantum efficiency
of Yb doped systems is higher, and therefore the majority of
commercial and scientific systems prefer to use Yb over Er. Most
systems operate at the $\lambda=1.03\mic$ Yb emission line and can for
low pulse energies ($<15~\mu$J) generate pulses as short as 250 fs,
while higher pulse energies result in longer pulses (currently
$50~\mu$J 450 fs pulses is the state-of-the-art for commercial
systems). In Er amplifier systems much lower pulse energies are
available, typically $1-3~\mu$J and $500-700$ fs pulses at
$\lambda=1.55\mic$; such low pulse energies and long pulse duration
mean that only very low soliton orders can be excited, and thus the
CQSC can only achieve very moderate compression occurring in very long
crystals. 

The basis for the following case studies and numerical simulations is
therefore 
a couple of commercially available Yb-based fCPA systems, both
operating at 1030 nm. Case (1) is a Clark MXR Impulse
\footnote{\htmladdnormallink{{h}ttp://www.clark-mxr.com}
{http://www.clark-mxr.com}} giving 15 $\mu$J 250 fs FWHM
pulses, which represents a system giving quite short, yet still
reasonably energetic pulses as a starting point. Case (2) is an
Amplitude Systemes Tangerine
\footnote{\htmladdnormallink{{h}ttp://www.amplitude-systemes.com}
{http://www.amplitude-systemes.com}} giving 50 $\mu$J 450 fs
FWHM pulses, which represents a system with more energetic but also
longer pulses.

\begin{figure}[tb]
  \includegraphics[width=8.5cm]{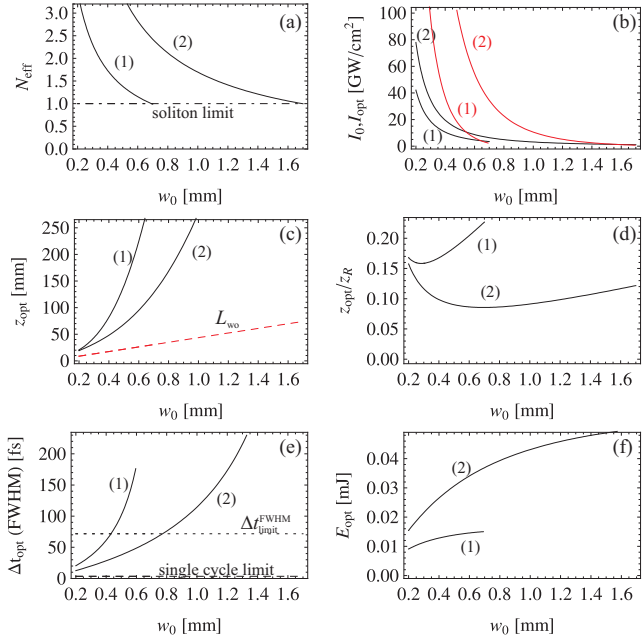}
\caption{\label{fig:or} (Color online) Practical operation range of
  the LN type I compression system at $\lambda_1=1.03\mic$ for $\Delta
  k=45\imm$. The plots show the predicted behaviour when the FW waist
  $w_0$ is varied. The two cases are (1) pump pulses with $\tin^{\rm
    FWHM}=250$ fs and 15 $\mu$J pulse energy, and (2) pump pulses with
  $\tin^{\rm FWHM}=450$ fs and 50 $\mu$J pulse energy. The curves in
  (b)-(f) are calculated based on $\neffs$ shown in (a) by using the
  scaling laws \cite{bache:2007} that hold for $\neffs>1$. }
\end{figure}

The two cases are studied together taking $\Delta k=45\imm$.
Figure~\ref{fig:or}(a) shows that in case (1) we need to focus the
pulses to $w_0<600\mic$ to observe solitons: in this regime
Fig.~\ref{fig:or}(d) shows that the Rayleigh length $z_R=\pi
w_0^2/\lambda$ is only 5-6 times larger than the optimal compression
point $z_{\rm opt}$ of around 100 mm. This is borderline at the risk
of experiencing diffraction problems. Even increasing or decreasing
the waist does not improve this ratio much. In case (2) instead, the
increased pulse energy makes solitons appear already at $w_0\simeq
1.6$ mm, despite the longer pulse duration. This means that
diffraction should be less of an issue: in Fig.~\ref{fig:or}(d) the
pulse compression point relative to the Rayleigh length of the focused
beam is significantly smaller in case (2).

Fig.~\ref{fig:or}(c) indicates that the spatial walk-off in the
crystal can become an issue: the crystal should be shorter than the
spatial walk-off length $L_{\rm wo}=w_0/\tan \rho\simeq w_0/\rho$ to
ensure proper interaction between the FW and the SH, but evidently
the pulse compression lengths in both cases are at least a factor of
2-3 longer than the spatial walk-off length. It might therefore be
necessary to compensate for this by using two crystals, one inverted
relative to the other so the walk-off direction in the 2. crystal is
inverted with respect to the 1. crystal \cite{Zondy:1994}. 

An alternative solution to the walk-off problem is to turn to a
noncritical phase matching scheme, where $\rho=0$. This happens for
$\theta=0$ or $\pi/2$, see Fig.~\ref{fig:1060nm}(d). Of course this
removes the possibility of tuning the phase matching via $\theta$, and
one has to turn to temperature tuning of $\Delta k$. 
The temperature needed to get to the desired operation point ($\Delta k\simeq
40-50\imm$) can be estimated using the temperature dependent Sellmeier
equations \cite{gayer:2008}, and our calculations indicate that it
should happen already at a temperature of around $45^\circ$ C. This
would make an easy solution to the walk-off problem. 

The strong GVM implies that compression of Yb-based systems can
only occur in the nonstationary regime, see Fig.~\ref{fig:LN-cw}.
Thus, unless $\neffs$ is close to unity the GVM induced Raman-like
effects dominate, and the FW pulse becomes extremely distorted and
very poorly compressed. Actually, as a rule of thumb it never makes
sense to use $\neffs$ larger than what is sufficient to reach the
limit expressed by Eq.~(\ref{eq:tlimit}), and typically even an
$\neffs$ smaller than that. The limit is drawn as a dotted line
in Fig.~\ref{fig:or}(e), and it is reached around $w_0=400\mic$ in
case (1) and $w_0=800\mic$ in case 2.

Finally, Fig.~\ref{fig:or}(b) shows that quite moderate input
intensities must be used to achieve solitons in both cases. This is
related to the quite long input pulse durations. Furthermore,
Fig.~\ref{fig:or}(f) shows that the low soliton orders conserve most
of the pulse energy in both cases.


\section{Numerical simulations}
\label{sec:Numer-simul}

We here present numerical simulations of the two cases using a
plane-wave temporal model based on the slowly evolving wave equation
(see more details in \cite{bache:2007} and references therein), which
includes self-steepening effects and higher-order dispersion. This
model is justified as long as diffraction is minimal, which we assume
is the case when the crystal length is much shorter than the Rayleigh
length, and when spatial walk-off is minimal. This requirement will be
discussed further below.

\subsection{Case (1): 250 fs 15 $\mu$J pulses}
\label{sec:Case-(1):-250}

\begin{figure}[tb]
  \includegraphics[width=8.5cm]{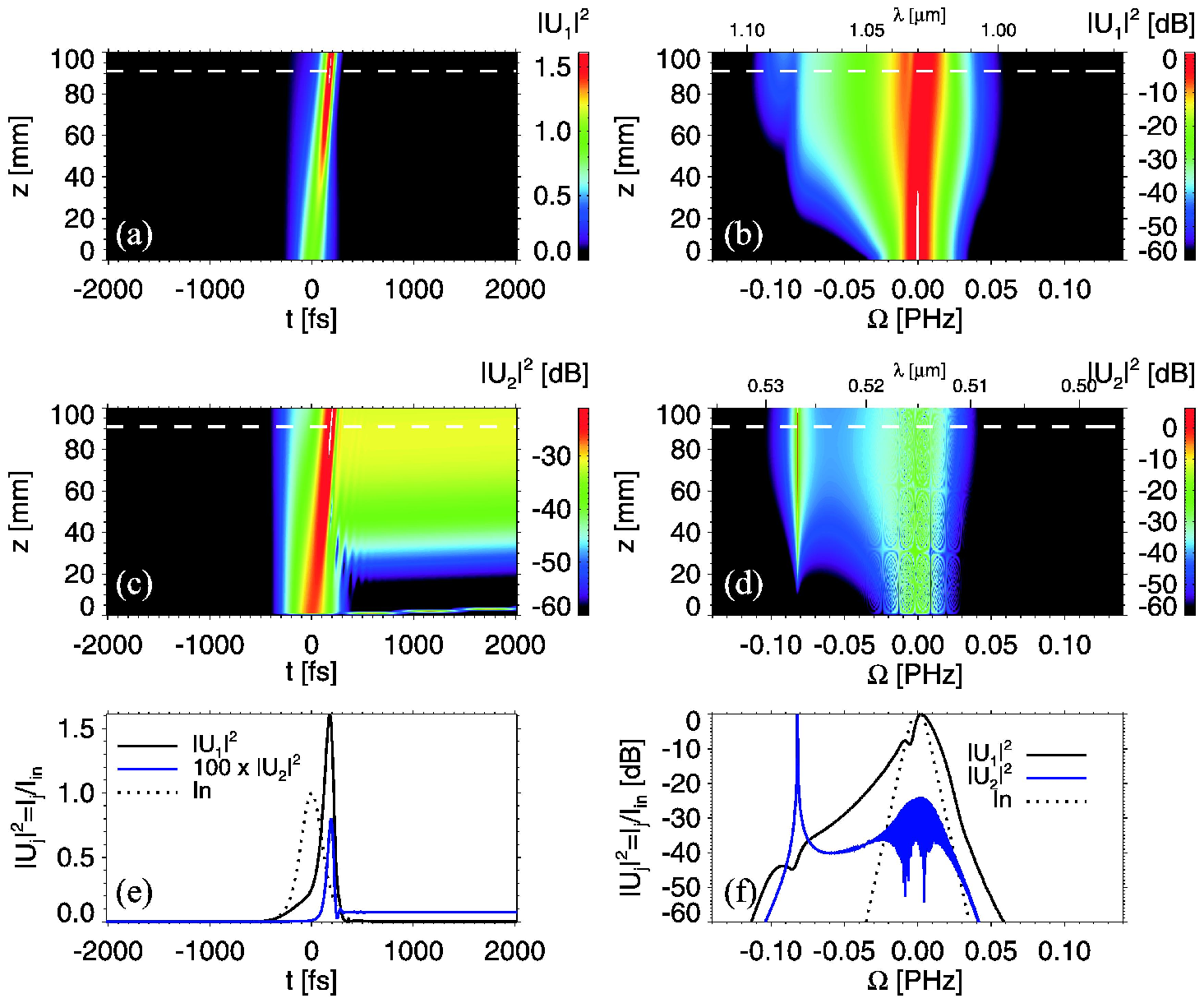}
\caption{\label{fig:compression2D} (Color online) Numerical simulation
  of soliton compression in LN with $\lambda_1=1.03\mic$, $\tin^{\rm
    FWHM}=250$ fs, $\Delta k=45\imm$ and $\neffs=1.4$ (implying
  $\Iin=6.9~{\rm GW/cm^2}$). The FW pulse shown in (a) compresses to
  $\topt=126$ fs (FWHM) after propagating 91 mm. The SH time plot (c)
  and FW (b) and SH (d) spectra are also shown on a logarithmic scale,
  and $U_j$ are normalized to the peak input FW electric field. In (e)
  and (f) cuts are shown at the optimal compression point $z=91$ mm
  (corresponding to the white line in the 2D plots). Note that the SH
  in (e) is magnified 100 times.}
\end{figure}

For the 250 fs $15~\mu$J pulses from a Clark laser system we found
that the best compression was obtained with $\neffs\sim 1.3-1.5$.  This
soliton order can be achieved with 15 $\mu$J pulse energy when the
pump is focused to around $w_0=400\mic$, see Fig.~\ref{fig:or}(a).

The theoretical compression factor for such soliton orders is
$f_c=2-3$, \ie \ a $\topt\sim 80-125$ fs FWHM compressed pulse is
predicted.  In Fig.~\ref{fig:compression2D} we show the results of a
simulation with $\neffs=1.4$. This soliton order gave the best
compression: a slightly asymmetric $\topt=126$ fs (FWHM) pulse is
observed after 91 mm of propagation, see (a) and cut in (e). The
compression is not quite as strong as predicted by the scaling
law~(\ref{eq:fc-fit}), but this is because the scaling laws are based
on pulse compression in the stationary regime. On the other hand the
pulse quality is large, $Q_c=0.82$, so most of the pulse energy is
retained in the central compressed part, and the pulse pedestal is also
very small. These  are the main advantages of soliton
compression with low soliton orders.

The FW spectrum (b) experiences upon propagation SPM-like broadening,
where the blue-shifted shoulder clearly dominates; this is a sign of
the cascaded quadratic nonlinearities dominating, and the fact that it
is blue shifted is related to the negative sign of $d_{12}$.

In the SH time-plot (c) we observe the strong GVM first inducing a
weak component quickly escaping from the central part of the pulse,
and later the GVM induces the characteristic DC-like trailing temporal
pulse in the SH (this often occurs close to or at phase matching in
presence of GVM, see also
\cite{Noordam:1990,bakker:1992,Su:2006}). 
This behaviour is also reflected in the SH spectrum, see (d) and cut in (f),
which shows a very strong and extremely narrow red-shifted component
building up, which eventually becomes the dominating contribution. As
we discuss below its spectral position can accurately be predicted by
the nonlocal theory that was recently developed by us
\cite{bache:2007a,bache:2008}. We believe
that this strong and long SH trailing component actually causes the trailing
part of the FW to be strongly depleted, and that this is the main
reason for the asymmetrical FW shape.

\begin{figure}[tb]
      \includegraphics[width=4.1cm]{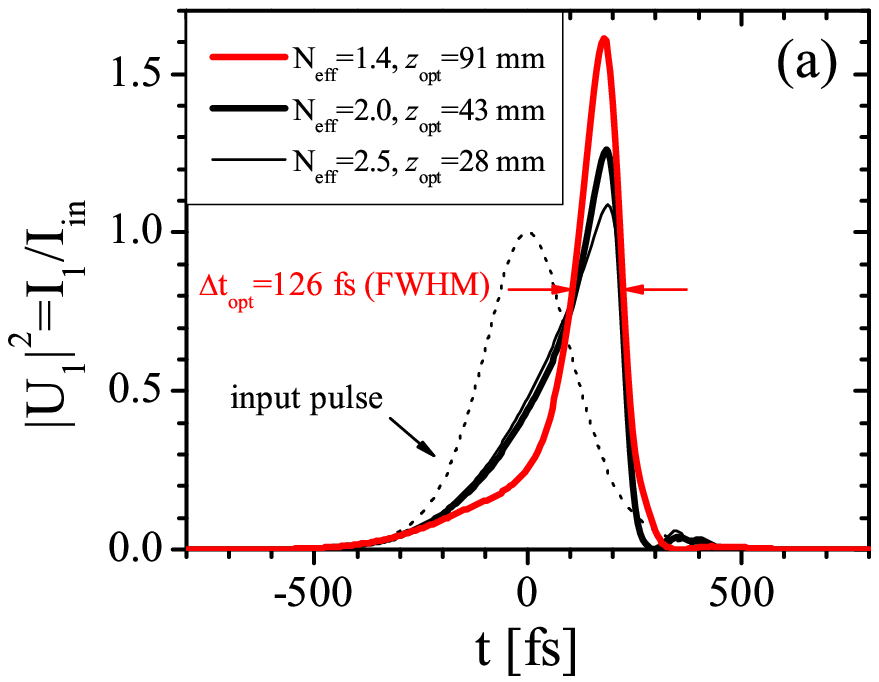}
      \includegraphics[width=4.3cm]{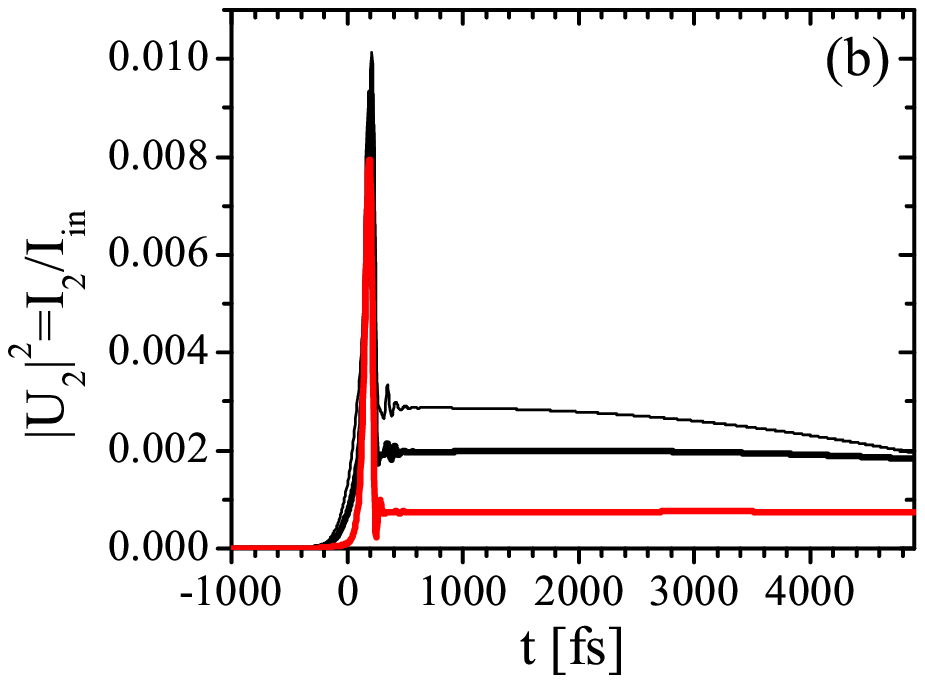}
\caption{\label{fig:compression-examples} (Color online) Simulations as in
  Fig.~\ref{fig:compression2D} but with increasing $\neffs$. The red
  curve corresponds to the optimal compression point from
  Fig.~\ref{fig:compression2D}(e), while the black curves show what
  happens as the effective soliton order increases (making the optimal
  compression point occurring sooner).}
\end{figure}

The question is now: can we increase the effective soliton order and
achieve further compression below 100 fs as to approach the limit
predicted by Eq.~(\ref{eq:tlimit})? This turns out to be impossible:
when $\neffs$ is increased the GVM effects become stronger, making the
compressed pulse more distorted. This is clearly observed in
Fig.~\ref{fig:compression-examples}, where we increase $\neffs$ and
compare with the compression of Fig.~\ref{fig:compression2D}: For
$\neffs=2.0$ the compressed FW pulse in (a) is still quite short, but
clearly is less clean. For $\neffs=2.5$ the compressed pulse instead
becomes quite distorted. It is also evident in the SH time plots that
the trailing DC-like component increases with $\neffs$, while the
central part in all cases is a sub-100 fs FWHM pulse.
It is quite weak because most of the converted SH energy is fed into 
the DC-like part of the pulse, which is connected to the strong
spectral peak in the SH spectrum. This spectral peak becomes
stronger with increased $\neffs$ (not shown), but does not change
position as it does not dependent on $\neffs$.

\begin{figure*}[thb]
    \centerline{
      \includegraphics[width=5.5cm]{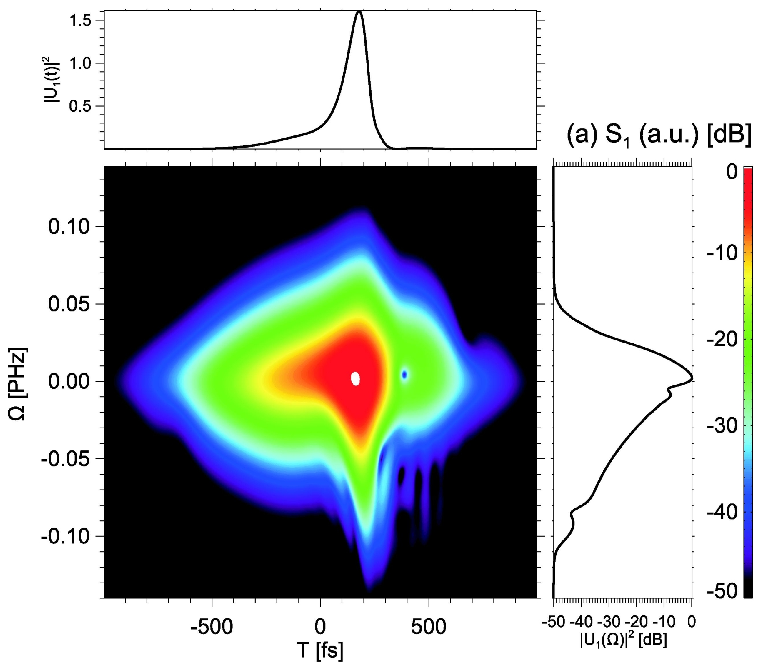}
      \includegraphics[width=5.5cm]{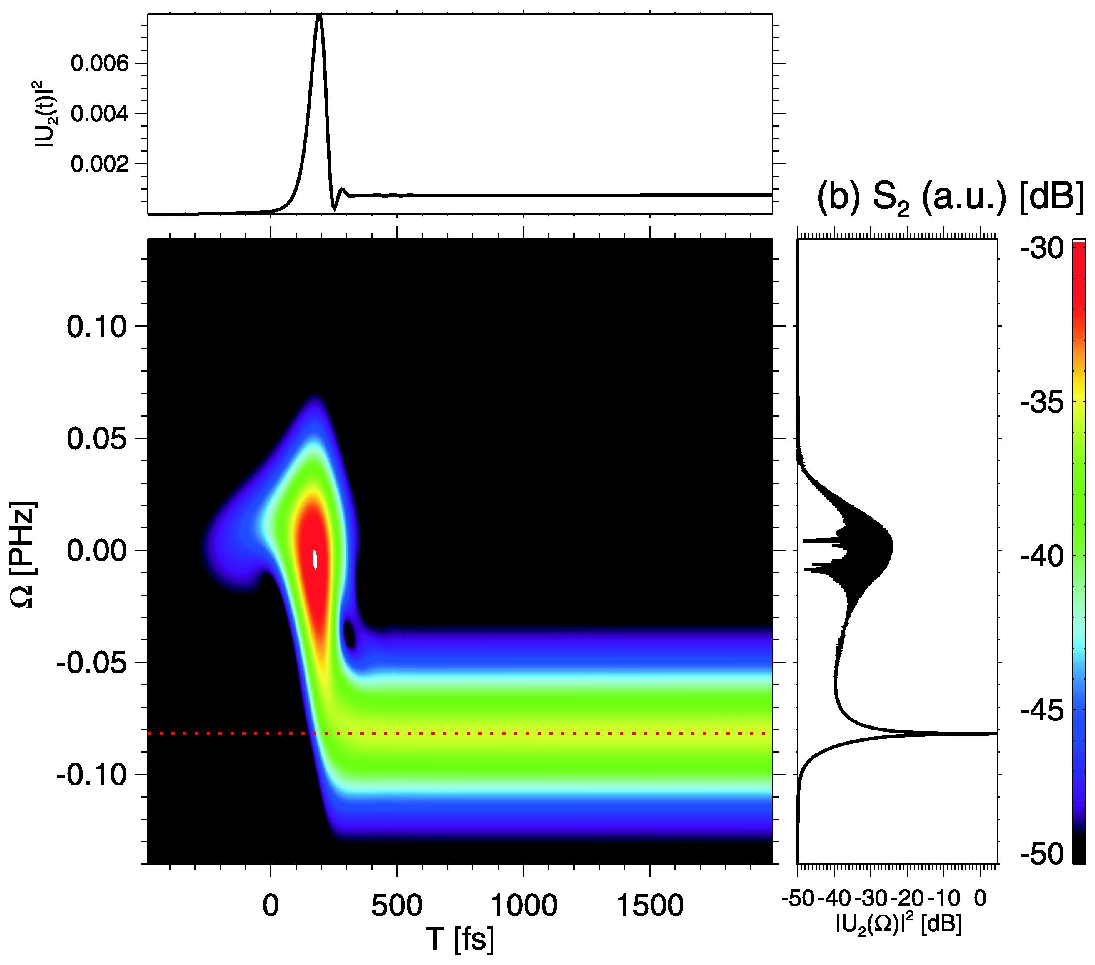}
      \includegraphics[width=5.5cm]{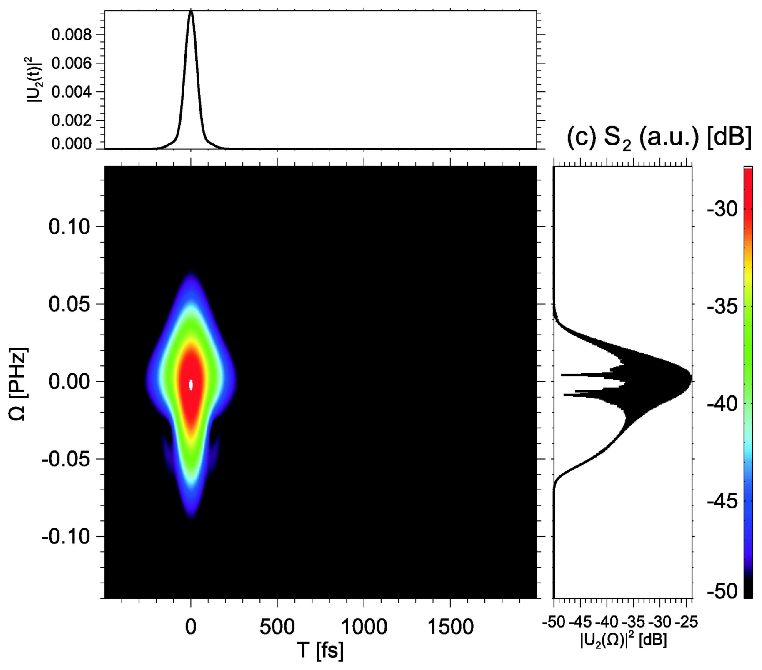}
    }
\caption{\label{fig:compression-spectrum} (Color online) XFROG-like
  spectrograms of the simulation in Fig.~\ref{fig:compression2D} at
  the optimal compression point $z_{\rm opt}=91$ mm. The sech-shaped
  gating pulse had $T_0^{\rm FWHM}=70$ fs, and the spectrograms are
  normalized to the peak value of $S_1$. The top and side plots show
  the purely temporal and spectral traces, respectively, and are thus
  identical to Fig.~\ref{fig:compression2D}(e) and (f). The red dashed
  line in (b) indicates the value $\Omega_+$ as calculated by the
  nonlocal theory.  The spectrogram in (c) shows the SH passed through
  a 3.  order super-Gaussian bandpass filter centered at the SH
  carrier frequency $\lambda_2=0.515\mic$ and with a FWHM of 100 THz.}
\end{figure*}

In order to understand the spectral content of the different temporal
components, the cross-correlation
frequency-resolved optical gating (XFROG) method is useful. The
spectral strength is given by
\cite{linden:1998}
\begin{eqnarray}
  \label{eq:xfrog}
  S_j(z,T,\Omega)=\left|\int_{-\infty}^{\infty}{\rm d} t 
  e^{i\Omega t}
  \Ef_j(z,t) \Ef_{\rm gate}(t-T) \right|^2
\end{eqnarray}
where $\Ef_{\rm gate}(t)$ is a properly chosen gating pulse. 
The spectrograms of the compressed pulses in
Fig.~\ref{fig:compression-examples} for $\neffs=1.4$ are shown in
Fig.~\ref{fig:compression-spectrum}. The FW compressed pulse is
slightly blue-shifted (around 2 THz), and the compressed part (located
at $T\sim 200$ fs) shows a significantly broader spectrum. 

The SH spectrum is very particular: the part of the pulse that
propagates with the FW group velocity (the ``locked'' part) shows a
quite clean short pulse. This group velocity locking of the SH has
been observed before
\cite{Noordam:1990,*bakker:1992,*Su:2006,ashihara:2004} and can be
understood from the nonlocal theory \cite{bache:2007a,bache:2008}: the
SH has a component that is basically slaved to the FW due to the
cascading nonlinearities. In frequency domain it can be compactly
expressed as \cite{bache:2008}
\begin{eqnarray}\label{eq:slave}
  U_2(z,\Omega)\propto \tilde R_-(\Omega) \FT[U_1^2(z,t)]
\end{eqnarray}
where $\FT[.]$ denotes the forward Fourier transform, and $U_j$ are
properly normalized fields. Thus, the spectral content of the SH is
slaved to the spectral content of the spectrum of $U_1^2$. The weight
is provided by the nonlocal Raman-like response function in the
nonstationary regime \cite{bache:2008}
\begin{eqnarray}\label{eq:response}
\tilde R_-(\Omega)=(2\pi)^{-1/2}\frac{\Omega_+\Omega_-}
{(\Omega-\Omega_-)(\Omega-\Omega_+)}  
\end{eqnarray}
where $\Omega_\pm=\Omega_a\pm \Omega_b$. These frequencies can be
calculated (to 2. order) from the dispersion of the system as
$\Omega_a=d_{12}/\kpp_2=-1.044$ PHz and $\Omega_b=|2\Delta
k/\kpp_2-\Omega_a^2|^{1/2}=0.963$ PHz. In the center around
$\Omega=0$, where $ \FT[U_1^2(z,t)]$ is residing in this case, the
response is quite flat: thus we get a SH component locked to the FW
and when the FW compresses so does this SH component.

Another striking feature of the SH spectrogram is the DC-like
component: it is very evident as a long pulse centered around
$\Omega\sim -80$ THz. Also this peak can be understood from
Eq.~(\ref{eq:slave}), because according to Eq.~(\ref{eq:response}) the
nonlocal response function in the nonstationary regime has sharp
resonance peaks in the response at $\Omega=\Omega_\pm$.  Inserting the
dispersion values of the simulation we get $\Omega_+=-81.6$ THz in
excellent correspondence with the observed peak position as the
red dashed line indicates.  Instead $\Omega_-$ is located too far
into the red side of the spectrum to affect the behaviour.

Considering this spectral composition, it might even be possible to
filter away the disturbing SH component at $\Omega=\Omega_+$, which in
time-domain would give a quite decent SH pulse. In (c) we show that
this is feasible: we pass the SH pulse through a super-Gaussian
($n=3$) bandpass filter centered at $\omega_2$ and with a bandwidth of 100
THz FWHM (corresponding to 15 nm): this filters away the disturbing sharp
peak, and a 80 fs FWHM pulse remains at $\lambda=0.515$ nm. The peak
intensity in this short pulse is around $0.006 \Iin=0.0414\gw$. If we
assume that it is created with 15 $\mu$J pulse energy focused to
$w_0=0.5$ mm to achieve $\neffs=1.4$, and that the generated SH has
roughly the same spot size, then the pulse energy of the filtered 80
fs pulse would be around 50 nJ.


\subsection{Case (2): 450 fs 50 $\mu$J pulses}
\label{sec:Case-(2):}

\begin{figure}[b]
      \includegraphics[width=4.1cm]{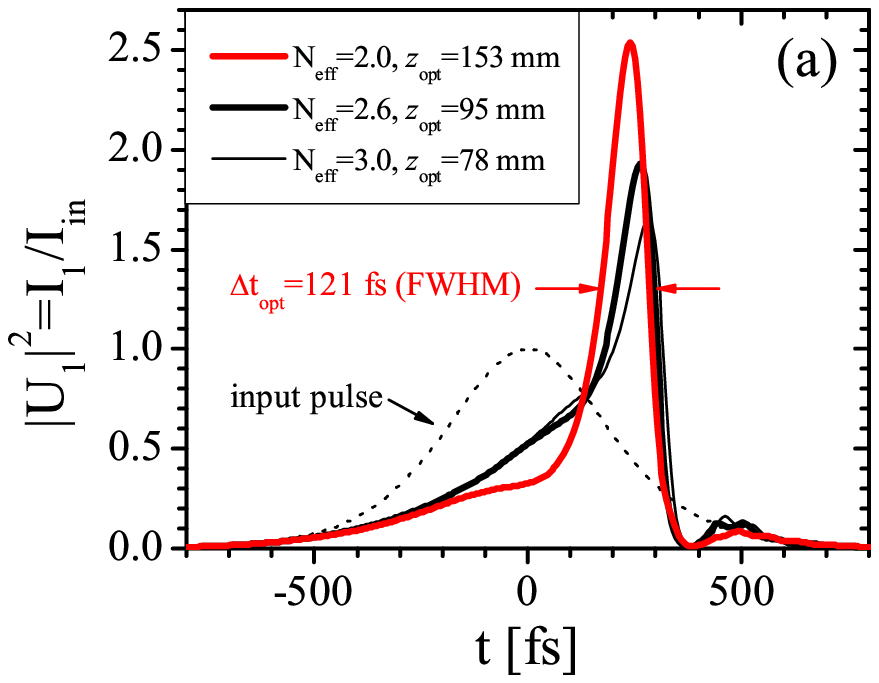}
      \includegraphics[width=4.3cm]{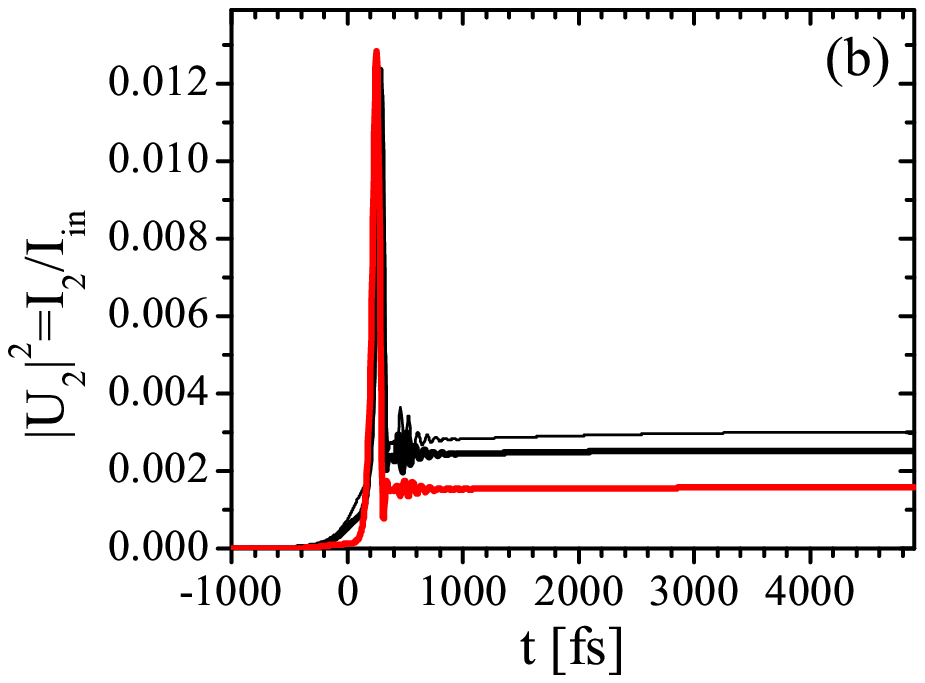}
\caption{\label{fig:compression-examples-450} (Color online) Numerical 
  simulations using 450 fs 50 $\mu$J input pulses and taking $\Delta
  k=45 \imm$. The best pulse was observed for $\neffs=2.0$ (red curve)
  where pulse compression occurs after 15 cm. The black curves show
  what happens as the effective soliton order increases (in which case
  the optimal compression point occurs sooner).}
\end{figure}

In case (2) the pulse duration is longer, 450 fs. When the pulse
duration is longer the soliton will for a fixed soliton order compress
after a longer distance.  This is because according to
Eq.~(\ref{eq:zopt-fit}) $\lmax\propto L_{\rm D,1}\propto \tin^2$.
However, we may compensate for this by increasing the effective
soliton order enough to reach the limit governed by
Eq.~(\ref{eq:tlimit}). For a 450 fs 50 $\mu$J pulse it is achieved
around $w_0=0.8$~mm, see Fig.~\ref{fig:or}(e), resulting in
$\neffs\sim 2.0-2.5$. This higher soliton order should make it
possible to compress in crystal lengths of around 10-15 cm, see
Fig.~\ref{fig:or}(c).

In Fig.~\ref{fig:compression-examples-450} we show some numerical
simulations using these longer more energetic pulses. The best pulse
observed shows a three-fold compression to $\topt=121$ fs (FWHM) at
$\neffs=2.0$. The compression occurred after around 15 cm propagation,
so spatial walk-off would be an issue here. Increasing the soliton
order to $\neffs=2.6$ the pulse becomes more distorted, but still
compresses to around 150 fs FWHM after 9.5 cm, a more realistic
interaction length. Finally, at $\neffs=3.0$ the pulse becomes too
distorted as the GVM effects become stronger.

In the two cases the pulses therefore eventually
compress to the same duration, which is the limit imposed by the
nonlocal GVM effects. The more energetic pulses in case (2) allow for
a more defocused pump beam so the compression should be less affected
by diffraction. On the other hand, as the pulses are longer they
compress later, so spatial walk-off is a more severe issue. A more
optimal situation in both cases would therefore be more energetic
pulses so the pump can be defocused with a factor 2-3. This would
diminish spatial walk-off effects.

\section{Conclusion}
\label{sec:Conclusion}

Here we have shown that lithium niobate (LN) crystals in a
type I cascaded SHG interaction can provide moderate compression of fs
pulses from Yb-based fiber amplifier systems ($1.03\mic$ wavelength).
The phase mismatch was controlled through angle tuning (critical phase
matching interaction). Using numerical simulations we found that the
best compression was to around 120 fs FWHM after around 10 cm propagation.

Better compression was prevented in part by strong GVM effects, caused
by strong dispersion in the LN crystal, and competing material Kerr
nonlinear effects. These are focusing of nature and counteract the
defocusing Kerr-like nonlinearities from the cascaded SHG. In order to
make the total nonlinear phase shift negative the phase mismatch had
to be taken quite low, and in this regime GVM effects dominate (the
``nonstationary'' regime). GVM imposes a strongly nonlocal temporal
response in the cascaded nonlinearity that feeds most of the converted
energy into a narrow red-shifted peak. In the temporal trace this gave
a SH with a multi-ps long trailing component. The FW therefore
experienced a distorted compression less the soliton order was kept
very low. For such low soliton orders the compression distance
increases substantially, but here the strong dispersion of the
LN crystal actually becomes an advantage: due to a large GVD the
soliton dynamics occur in much shorter crystals than usual, and
the numerics indicated compression in realistic crystal lengths (10
cm). 

It was noted that using low soliton orders gave a
compressed pulse retaining most of the input pulse energy (in the cases
we showed around 80\%), and that the unavoidable soliton pedistal was
less pronounced.

We also discussed the implications of using long crystals. Spatial
walk-off will be an issue since it is a critical phase matching scheme
is used that exploits birefringence, and also diffraction can be a problem. In
order to counteract these detrimental effects the pump pulses need to
be as energetic and short as possible. Two cases were highlighted
taken from commercially available systems, and we argued that
diffraction should not prevent observing the predicted compression,
but that some sort of walk-off compensation might be needed. Future
systems with more energetic pulses and reasonably short pulse
durations ($<500$ fs) would be able to beat the walk-off problem.
Walk-off could also be prevented by using a noncritical type I phase
matching scheme ($\theta=\pi/2$) and increasing the temperature
slightly to around $45^\circ$ C.

We finally noted that the peculiar SH shape in the nonstationary
regime gave a very characteristic spectrogram: as mentioned above
nonlocal GVM effects resulted in a sharp spectral red-shifted peak
with a long multi-ps trailing temporal component. Another pulse
component was instead locked to the group velocity of the compressed
FW soliton. This locked visible pulse was located at the SH wavelength
(515 nm), quite far from the red-shifted peak. We showed that a simple
bandpass filter could actually remove the detrimental red-shifted peak
leaving a very clean 80 fs visible pulse ($\lambda=515$ nm).  This is
the opposite approach compared to other studies, see e.g.
\cite{bakker:1992,Marangoni:2007}, where focus was on exploiting
``spectral compression'' of fs pulses to obtain longer ps pulses.
Despite that the cascaded SHG by nature has a low conversion
efficiency, the pulse energy of this short visible pulse can easily be
50-100 nJ. Such pulses could be used for two-color ultra-fast
energetic pump-probe
spectroscopy.

This study showed that cascaded quadratic pulse compression is
possible even in a very dispersive nonlinear crystal. However, if
compression occurs in a medium with stronger quadratic nonlinearities
then it would be possible to increase the phase mismatch, and thereby
enter the stationary regime where the nonlocal GVM effects are much
weaker. The benefit would be triple: cleaner compressed pulses could
be generated, higher soliton orders could be used to achieve stronger
compression, and it would occur in a shorter crystal.
This conclusion is in line with what was noted previously in a fiber
context \cite{Bache:2009}, where one of us found that the very
dispersive nature of wave-guided cascaded SHG could be overcome if a
strong enough quadratic nonlinearity is present. We are currently
investigating other possible nonlinear crystals and phase matching
conditions to achieve this.

\section{Acknowledgments}

\label{sec:Acknowledgements}

Support is acknowledged from the Danish Council for Independent
Research (Technology and Production Sciences, grant no. 274-08-0479
\htmladdnormallink{\textit{Femto-VINIR}}{http://www.femto-vinir.fotonik.dtu.dk},
and Natural Sciences, grant no. 21-04-0506).  Jeff Moses and Binbin
Zhou are acknowledged for useful discussions.

\appendix

\section{LN crystal parameters}
\label{sec:Crystal-parameters}

LN is a negative uniaxial crystal of symmetry class $3m$.  Its low
damage threshold due to photorefractive effects and problems with green
induced IR absorption can be improved
dramatically by doping the crystal, in particular with MgO doping
\cite{bryan:1984,*furukawa:2000,*Furukawa:1998,furukawa:2001}.
1\% MgO doping in stoichiometric LN (1\% MgO:sLN) is enough to
practically remove photorefractive effects and increase
dramatically the damage threshold, 
while 5\% is needed in congruent LN (5\% MgO:cLN) to do the same
\cite{furukawa:2001}. 1\% MgO:sLN also has a shorter UV absorption
edge ($\lambda=0.31\mic$).

We here use 1\% MgO:sLN, and the Sellmeier equations from 
\cite{gayer:2008}: note that for 1\% MgO:sLN they only measured $n_e$,
but we checked that the 5\% MgO:cLN $n_o$ Sellmeier equation matches
well (at room temperature) the 1\% sLN $n_o$ equation from
\cite{nakamura:2002}. The quadratic nonlinear coefficients have been
measured at $\lambda=1.06\mic$ and are $d_{31}=-4.7$ pm/V and
$d_{33}=23.8$ pm/V \cite{Shoji:2007}, while $d_{22}=2.1$ pm/V
\cite{miller:1971,*dmitriev:1999} was measured for undoped LN. The
fact that $d_{31}d_{22}<0$ has been established in, e.g.,
\cite{klein:2003a}. The effective quadratic nonlinearity of the type I
$oo\rightarrow e$ interaction is given by Eq.~(\ref{eq:deff}).
Because $d_{31}d_{22}<0$ \cite{klein:2003a} the maximum nonlinearity is 
realized with $\phi=-\pi/2$. 

\section{Anisotropic Kerr nonlinear refraction}
\label{sec:Anis-Kerr-nonl}

We previously studied type I cascaded SHG in a BBO crystal
\cite{bache:2007a,bache:2007,bache:2008}, assuming an
isotropic Kerr nonlinearity 
\begin{eqnarray}
\chi_{\rm eff,11}^{(3)}=\chi_{\rm eff,22}^{(3)}=3\chi_{\rm eff,12}^{(3)}  
\end{eqnarray}
where $\chi_{{\rm eff},jj}^{(3)}$ are the FW and SH SPM coefficients,
and $\chi_{\rm eff,12}^{(3)}$ is the XPM coefficient. However, all
quadratic nonlinear crystals are anisotropic, and below we address
this. 

\begin{figure}[tb]
\includegraphics[width=3cm]{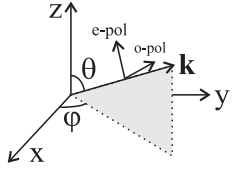}
\caption{\label{fig:crystal} Definition (in accordance with
  the IRE/IEEE standard \cite{roberts:1992}) of the crystal coordinate
  system $xyz$ relative to the beam propagation direction indicated by
  ${\mathbf k}$.}
\end{figure}

Note first that the error made in assuming an isotropic response for
the CQSC is probably small as the crucial parameter is the FW SPM
coefficient. As we will see now for type I this is identical in the
isotropic and in the anisotropic cases. However, it should be
emphasized that the various experimental attempts to measure the Kerr
nonlinear refractive index of nonlinear crystals do not always measure
the tensor component relevant to our purpose, namely the $c_{11}$
component, see Table~\ref{tab:ln} later. The analysis presented here
should help understanding what exactly has been measured, and put the
results into the context of cascaded quadratic soliton compression.

For a nonlinear crystal in the symmetry group $3m$ (LN and BBO) there are
37 nonzero elements for the $\underline{\underline\chi}^{(3)}$ tensor,
and of these only 14 are independent \cite{boulanger:2006}
\begin{eqnarray}
  xxxx &=& yyyy = xxyy + xyxy + xyyx\nonumber\\
 xxzz &=& xzxz = xzzx = yyzz = yzyz = yzzy \nonumber\\
&=& zyyz = zyzy = zzyy = zxxz =
 zxzx = zzxx\nonumber \\
 xxyy &=& xyxy = xyyx = yxxy = yxyx = yyxx\nonumber\\
 xxyz &=& xxzy =
 xyxz = xyzx = xzxy = xzyx\nonumber\\
&=&-yyyz=-yyzy=-yzyy 
= yxxz = yxzx \nonumber\\
&=& yzxx = -zyyy = zxxy = zxyx = zyxx\nonumber \\ zzzz&
\end{eqnarray}
where Kleinman symmetry has been invoked, and the polarization
relative to the crystal coordinate
system is defined in Fig.~\ref{fig:crystal}. Under Kleinman symmetry the
nonlinear coefficients are assumed dispersionless and the criterion
for this assumption is that the system is far from any resonances.
Using the notation $\chi_{ijkl}^{(3)}=c_{\mu m}$ where
\begin{eqnarray}
  {\rm for~\mu:} &\quad x \rightarrow 1\quad y \rightarrow 2
  \quad z \rightarrow 3\nonumber\\
  {\rm for~}m: &\quad xxx \rightarrow 1 \quad yyy \rightarrow 2
  \quad  zzz \rightarrow 3 \quad yzz \rightarrow 4 \nonumber\\
  & \quad yyz \rightarrow 5 \quad  xzz \rightarrow 6 \quad xxz \rightarrow 7
  \quad xyy \rightarrow 8 \nonumber\\
&\quad xxy \rightarrow 9 \quad xyz
  \rightarrow 0
\end{eqnarray}
these tensor components are equivalent to
\begin{eqnarray}\label{eq:c-Kleinman}
  c_{11} &=& c_{22} = 3c_{18} \nonumber\\ 
 c_{16} &=& c_{24}= c_{35} = c_{37}\nonumber \\
 c_{18} &=& c_{29}\nonumber\\
 c_{10}&=&-c_{25} = c_{27} =
 -c_{32} = c_{39} \nonumber\\
 c_{33}&
\end{eqnarray}
On the reduced form the cubic tensor becomes
\begin{eqnarray}\label{eq:chi3}
&&\underline {\underline c}=\\
  &&\begin{bmatrix}
c_{11}     & 0      & 0      & 0     & 0       & c_{16} &  0      & \tfrac{c_{11}}{3} & 0 & c_{10} \\
0 & c_{11} & 0      & c_{16}& -c_{10} & 0      &  c_{10} & 0        & \tfrac{c_{11}}{3} & 0 \\
0 & -c_{10}& c_{33} & 0     & c_{16} & 0       & c_{16}  &  0      & c_{10} & 0 \\
  \end{bmatrix}
\nonumber
\end{eqnarray}
These results conform with the IRE/IEEE standard \cite{Banks:2002}.


We now want to evaluate the cubic nonlinear response for a type I
interaction. Using the notation from
\cite{bache:2007} the cubic nonlinear polarization response is
\begin{eqnarray}
  \label{eq:pol}
  \Pol_{\rm NL}^{(3)}=\evac
\underline{\underline{\chi}}^{(3)}\vdots
  \E \E \E 
\end{eqnarray}
We have here only considered an instantaneous (electronic) cubic
nonlinear response \cite{[{The Raman response of LN has been studied
    in the past, see, e.g., }][{, but since we deal with quite long
    pulses $>50$ fs it is safe to neglect such effects in our
    simulations}]barker:1967}. Let us consider the type I SHG
interaction where two ordinarily polarized FW photons are converted to
an extraordinarily polarized SH photon ($oo\rightarrow e$). In the
coordinate system according to the IRE/IEEE standard \cite{roberts:1992},
see Fig.~\ref{fig:crystal}, the unit vectors for $o$-polarized and
$e$-polarized light are
\begin{eqnarray}\label{eq:ev-typeI}
  \ev^o=
  \begin{bmatrix}
    -\sin\phi\\\cos\phi\\0
  \end{bmatrix}
\quad
  \ev^e=
  \begin{bmatrix}
    -\cos\theta\cos\phi\\-\cos\theta\sin\phi\\\sin\theta
  \end{bmatrix}
\end{eqnarray}
where walk-off has been neglected. 

We then introduce
slowly varying envelopes polarized along arbitrary directions
\begin{eqnarray}\label{eq:e-svea}
\E(t)={\rm Re}[\uv_1 \Ef_1(t)e^{-i\omega_1 t}+ \uv_2\Ef_2(t)e^{-i\omega_2 t} ]  
\end{eqnarray}
where $\uv_j$ is the unit polarization vector. For type I SHG
we have $\uv_1=\ev^o$ and $\uv_2=\ev^e$. The nonlinear slowly varying
polarization response 
\begin{eqnarray}\nonumber
 \Pol_{\rm NL}^{(3)}(t)={\rm Re}[\uv_1 P_{\rm NL,1}^{(3)}(t)e^{-i\omega_1 t}+
\uv_2P_{\rm NL,2}^{(3)}(t)e^{-i\omega_2 t} ] 
\end{eqnarray}
then becomes
\begin{eqnarray}\label{eq:pol-typeI-svea}
  P_{{\rm NL},i}^{(3)}=\frac{3}{4}\evac \left[\chi^{(3)}_{{\rm
  eff},ii}|\Ef_i|^2+
  2\chi^{(3)}_{{\rm eff},ij}|\Ef_{j}|^2\right]\Ef_i, 
\end{eqnarray}
where $i,j=1,2$ and $j\neq i$. We have here only included
phase-matched components and frequency-mixing terms where $2\omega_1
-\omega_2=0$. The numerical prefactor $\tfrac{3}{4}$ is the $K$-factor
\cite{butcher:1990} for a third order nonlinear effect creating an
intensity dependent refractive index with degenerate frequencies, and
the factor 2 on the XPM terms $\chi^{(3)}_{{\rm eff},ij}$ stems from
the fact that the $K$-factor for cross-phase modulation with
non-degenerate frequencies is
$\tfrac{3}{2}$.  

For calculating the cubic nonlinear coefficients, it is
convenient to use an effective cubic
nonlinearity \cite{Yang:1995}
\begin{eqnarray}\label{eq:chi3-eff}
  \chi_{\rm eff}^{(3)}=\uv_d\cdot
  \underline{\underline{\chi}}^{(3)}\vdots
  \uv_a\uv_b\uv_c=\uv_d\cdot\underline{\underline{c}}\cdot \uv^{(3)}, 
\end{eqnarray}
$ a,b,c,d=1,2$. Here $\uv_d$ is the unit vector of the field under
consideration; thus, if we are interested in calculating the cubic
nonlinear polarization for the FW [taking $i=1$ in
Eq.~(\ref{eq:pol-typeI-svea})], then $\uv_d=\uv_1$. The other three
unit vectors $\uv_{a,b,c}$ are the unit vectors of each field
appearing in Eq.~(\ref{eq:pol}), and can in the case we are
considering here be either $\uv_1$ or $\uv_2$ according to the
identity~(\ref{eq:e-svea}). Most combinations are
not phase matched or have $2\omega_1 -\omega_2\neq 0$, and are
therefore not included in Eq.~(\ref{eq:pol-typeI-svea}). 
The rank 4
tensor on reduced form, as given by Eq.~(\ref{eq:chi3}) for LN, can be
used to find the tensor product
$\underline{\underline{\chi}}^{(3)}\vdots \uv_a\uv_b\uv_c$ as a simple
matrix-vector product $\underline{\underline{c}}\cdot \uv^{(3)}$ where
\begin{eqnarray}\label{eq:uv3}
  \uv^{(3)}=
   \begin{bmatrix}
     L_{xxx}\\
     L_{yyy}\\
     L_{zzz}\\
     L_{yzz}+L_{zyz}+L_{zzy}\\
     L_{yyz}+L_{yzy}+L_{zyy}\\
     L_{xzz}+L_{zxz}+L_{zzx}\\
     L_{xxz}+L_{xzx}+L_{zxx}\\
     L_{xyy}+L_{yxy}+L_{yyx}\\
     L_{xxy}+L_{xyx}+L_{yxx}\\
     L_{xyz}+L_{xzy}+L_{zxy}+L_{yxz}+L_{yzx}+L_{zyx}\\
   \end{bmatrix}
\end{eqnarray}
Here $L_{jkl}\equiv u_{a,j}u_{b,k}u_{c,l}$ where the $jkl$
indices refer to the $x$, $y$ or $z$ components of the unit
vectors. 

It is convenient at this stage to simplify the notation based on the
type I SHG interaction we are interested in. The effective cubic
nonlinearity~(\ref{eq:chi3-eff}) then reduces to the nonlinear
coefficients appearing in Eq.~(\ref{eq:pol-typeI-svea}) 
\begin{eqnarray}\label{eq:chi3-eff-ij}
  \chi^{(3)}_{{\rm eff},ij}=\uv_i\cdot
  \underline{\underline{\chi}}^{(3)}\vdots
  \uv_i\uv_j\uv_j
\end{eqnarray}

The SPM terms can now be calculated as follows. The FW SPM
interaction has $i=j=1$ in Eq.~(\ref{eq:chi3-eff-ij}), and is an
$ooo\rightarrow o$ process: $\uv_1=\ev^o$.  
The SH SPM interaction has $i=j=2$ and is an $eee\rightarrow e$
process, so $\uv_2=\ev^e$. We then need to calculate
$\underline{\underline{\chi}}^{(3)}\vdots \uv_i\uv_i\uv_i$ using the
reduced notation. Since for the SPM terms all the unit vectors in $\uv^{(3)}$ are degenerate
in frequency, all $L_{jkl}$ components in a given vector
entry are identical, e.g.  $L_{yzz}=L_{zyz}=L_{zzy}$.  We then  get
for the FW
\begin{eqnarray}
  \underline{\underline{\chi}}^{(3)}\vdots \uv_1\uv_1\uv_1=
  \begin{bmatrix}
    -c_{11}\sin\phi \\  c_{11}\cos\phi  \\ -c_{10}\cos^3\phi
  \end{bmatrix}
\end{eqnarray}
A similar expression can be calculated for the SH SPM component,
although it is substantially more complex. In the final step we carry
out the vector dot product of these vectors with $\uv_i$, as dictated
by Eq.~(\ref{eq:chi3-eff-ij}), and get for the FW ($i=1$) and the SH
($i=2$) \cite{Banks:2002}
\begin{eqnarray}\label{eq:chi3-SPM-FW}
  \chi^{(3)}_{{\rm eff},11}&=&c_{11}\\
  \chi^{(3)}_{{\rm
  eff},22}&=&-4c_{10}\sin\theta\cos^3\theta\sin 3\phi+
  c_{11}\cos^4\theta \nonumber \\
\label{eq:chi3-SPM-SH}
&&+\tfrac{3}{2}c_{16}\sin^22\theta+c_{33}\sin^4\theta
\end{eqnarray}

For the XPM
terms note that the three unit vectors used to calculate
Eq.~(\ref{eq:uv3}) are non-degenerate in frequency.  As an example,
for $\underline{\underline{\chi}}^{(3)}\vdots \uv_2\uv_1\uv_1$ terms
like $L_{xyy}+L_{yxy}+L_{yyx}$ must be evaluated, whose components are
$L_{xyy}=-\cos\theta \cos^3\phi$ and
$L_{yxy}=L_{yyx}=\cos\theta\sin^2\phi\cos\phi$. This gives
$\chi^{(3)}_{{\rm eff},12}=\chi^{(3)}_{{\rm eff},21}$
and \cite{Banks:2002,Kulagin:2006}
\begin{eqnarray}
\label{eq:chi3-XPM}
  \chi^{(3)}_{{\rm
  eff},12}=&\tfrac{1}{3}c_{11}\cos^2\theta
+c_{16}\sin^2\theta +c_{10}\sin 2\theta\sin 3\phi
\end{eqnarray}

\begin{figure}[tb]
  \centerline{
\includegraphics[width=6cm]{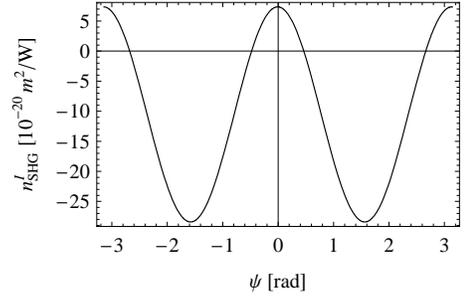}
  }
\caption{\label{fig:kulagin} The calculated effective Kerr nonlinear
  contributions from cascaded SHG to the measured Kerr nonlinear
  refractive index by Kulagin et al. \cite{Kulagin:2006}. The beam
  propagates with $\theta=\pi/2$ into a LN crystal, and $\psi$ denotes
  the polarization angle ($\psi=0$ gives $o$-polarized light, while
  $\psi=\pi/2$ gives $e$-polarized light). The angle $\phi$ was not
  reported, but we checked it has little influence on the $\ns$ value
  shown here.}
\end{figure}

\begin{table*}[tb]
  \centering
  \begin{tabular}
{
c          |c             |c         
|c                |c           |c    |c     |c     |c| c        |c|l}
$\lambda$ & $\chi^{(3)}_{\rm eff}$ & $n_{\rm Kerr}^I$ &
$n_{\rm Kerr}^I$  & $t_{\rm FWHM}$ & Rep. & $\theta$ & pol & $n $ & $c_{ij}$
& Ref. & Note  \\
$[{\rm nm}]$ & $[10^{-13}~{\rm esu}]$ & $[10^{-13}~{\rm esu}]$ &
    [$10^{-20}~{\rm m^2/W}$] & [ps] &  & [deg] & & &  &  \\ 
\hline
1064 & 1.1 & \underline{4.8} & 9.1  & 30 & single & 90 & $e$& 2.2 &
$c_{33}$ & \cite{desalvo:1996} & $x$-cut\\
1064 & 0.73 & \underline{3.2}  & 6.0  & 55 & 2 Hz & 90 & $o$& $2.2337\footnote{Linear refractive index not provided; this value was calculated
  by us for conversion purposes. }$ &
$c_{11}$ & \cite{gannev:2004} & paraxial fit\\
1064 & 0.66 & \underline{2.9}  & 5.4  & 55 & 2 Hz & 90 & $o$& $2.2337^{\rm a}$
&$c_{11}$ & \cite{gannev:2004} & Gaussian fit\\
1064 & \underline{2.4}  & 10 & 19 & 55 & 2 Hz & 90 &$o$  & 2.2337 &
$c_{11}$ & \cite{Kulagin:2006} & Fit to transmission curve\\
1064 & 0.80 & 3.4  & 6.3  & 55 & 2 Hz & 90 & $e$+$o$  & 2.2337 &
$c_{12},c_{18}$ & \cite{Kulagin:2006} &  
'', $c_{12}=c_{11}/3$\\ 
1064 & 0.57 & 2.8  & 4.9  & 55 & 2 Hz & 90  & $e$ & 2.1495 &
$c_{33}$ & \cite{Kulagin:2006} & '', $c_{33}=c_{12}/1.4$\\
1064 & 0.67 & 2.9  & 5.5  & 55 & 2 Hz & 90 & $e$+$o$ & 2.1912 & $c_{23},c_{16}$ &
\cite{Kulagin:2006} & '', $c_{23}=c_{12}/1.2$\\
800  & 1.8  & 7.8  & \underline{15} & 0.42 & 1 kHz & ? & ? & $2.1677^{\rm a}$ & ? &
\cite{burghoff:2007} & $x$-cut, $z$-cut \\
780  & 2.6  & 11.0 & \underline{20} & 0.15 & 76 MHz & 0 & $o$ & $2.2552^{\rm a}$ & $c_{11}$ &
\cite{Li:2001} & 6\% MgO:LN, $z$-cut \\
577  & \underline{1.6} & 6.6 & 12 &  5,000 & 40 Hz & 0 & $o$ & $2.301^{\rm a}$ & $c_{18}$ &
\cite{wynne:1972} & $c_{18}=c_{11}/3$ \\
532 & 10  & \underline{44} & 83 & 22   & single & 90 & $e$ & 2.23 & $c_{33}$ &
\cite{desalvo:1996} & $x$-cut\\
532 & 6.6 & 28 & \underline{53} & 25  &  10 Hz & 0  & $o$ & $2.2244^{\rm a}$ & $c_{11}$ &
\cite{li:1997} & $z$-cut\\
520 & \underline{5.0} & 21 & 39 & 0.2 & 1 kHz & 90 & $e$ & 2.24 & $c_{33}$ &
\cite{wang:2005} & 5\% MgO 0.06\% Fe cLN
  \end{tabular}
  \caption{Nonlinear Kerr refractive index of LN measured mainly by the Z-scan
    method \cite{sheik-bahae:1990}. The underlined results are the
  values reported. The 
  other entries have been calculated using
  Eqs.~(\ref{eq:n2I-n2-esu})-(\ref{eq:n2-I-esu-SI}). 
}
  \label{tab:ln}
\end{table*}

The next step is to obtain the the values for LN of each component in
Eqs.~(\ref{eq:chi3-SPM-FW})-(\ref{eq:chi3-XPM}). 
The value of the cubic nonlinear refractive index has been measured by
many authors and for many different pulse durations and crystal cuts.
In Tab.~\ref{tab:ln} the $\chi^{(3)}$ tensor components and the
$n_{\rm Kerr}^I$ are reported in electrostatic units values, and the
latter is also given in SI units (see
App.~\ref{sec:Conversion-relations} for details).

In one of the earliest studies the tensorial nature of LN was
studied \cite{wynne:1972}. Another early study found that
$c_{11}=3c_{10}$ \cite{eichler:1977}. Later studies used Z-scan
methods and often a nonlinear refractive index value was found without
any mentioning of the tensorial nature of the cubic nonlinear
susceptibility. The cascaded quadratic contributions were also often
forgotten or neglected. 

A recent study by Kulagin et al. went into a detailed experimental
determination of the various cubic tensor components of LN, and found
$c_{11}=2.4\times 10^{-13}$ esu at $\lambda=1.06\mic$, and 
that $c_{18}=1.2c_{16}=1.4 c_{33}$ \cite{Kulagin:2006}.  Through the
relation $c_{11}=3c_{18}$, see Eq.~(\ref{eq:c-Kleinman}), the other
coefficients are $c_{16}=c_{11}/3.6$,
$c_{33}=c_{11}/4.2$. A problem with this study is that the cascaded
quadratic nonlinear contributions to the observed Z-scan results were
neglected. Instead, based on an analysis of the anisotropic Kerr
tensor components the Z-scan transmission
function was calculated, and the various tensor components were found
by fitting to experimental data.  In the experiment the pump
propagated with $\theta=\pi/2$, i.e.  with the $\mathbf k$-vector
perpendicular to the OA. The angle of the polarization vector was then
varied; this gives either pure $o$-polarized light, pure $e$-polarized
light, or a linear mixture.

We have done an analysis of the various cascaded SHG processes that
come into play ($oo\rightarrow o$, $oo\rightarrow e$, $oe\rightarrow
e$, $oe\rightarrow o$, $ee\rightarrow e$, and $ee\rightarrow o$),
evaluated their respective $\deff$-values and phase mismatch values as
the input polarization angle changes. In total we arrived at a
strongly varying cascaded contribution shown in
Fig.~\ref{fig:kulagin}. At $\psi=0$ the contribution from $\ns$ is
focusing, implying that the $c_{11}$ component in Kulagin et al. might
be too high with a factor of $7.0\times 10^{-20}~\rm m^2/W$. There are
also strongly defocusing contributions at other polarization angles,
which should give rise to an underestimated value of the other tensor
components. Moreover, the overall shape reminds strongly of the shape
found in Fig. 5 in \cite{Kulagin:2006}: the focusing peaks from
cascaded SHG could explain the valleys found there, and the defocusing
valleys from cascaded SHG could instead explain the peaks. In summary
we believe the $c_{11}$ value to be too high, and the relation to the
other tensor components to be dubious.

There are other issues with the Z-scan method: If the repetition rate
is too high, there will also be contributions to the measured $\nk^I$
from thermal effects as well as two-photon excited free carriers
\cite{krauss:1994}, and hence $\nk^I$ does not contain
just the instantaneous electronic response, as it is supposed to. Similarly
conclusions can be made for pulses longer than 1 ps. For more on these
issues, see e.g.  \cite{Gnoli:2005}.

For the CQSC system the by far most important component is the FW SPM
coefficient $n_{\rm Kerr,11}^I$. The SH SPM and the XPM coefficients
only play minor roles in extreme cases close to transitions (e.g.,
close to the soliton existence line in Fig.~\ref{fig:LN-cw}). We
checked in the cases we studied in this paper that even increasing the
SH SPM and XPM Kerr coefficient several times the isotropic values did
not significantly change the compression results. 

Therefore until detailed reliable measurements of the cubic tensorial
components of LN become available, we decided to use an isotropic Kerr
response, and focus on using a realistic value of the FW SPM
coefficient. The best choice seems to be $\nk^I=20\times 10^{-20}~\rm
m^2/W$ at $\lambda=0.78\mic$ found in Ref. \cite{Li:2001}.  In this
experiment they have $\theta=0$ and thus what they measure is
$\chi^{(3)}_{\rm eff}=c_{11}$.  For orthogonal input polarization
(corresponding to $\phi=0,\pi/2$, both cases $o$-polarized) they find
the same value as they should since this $\chi^{(3)}_{\rm eff}$ does
not depend on $\phi$, cf. Eq.~(\ref{eq:chi3-SPM-FW}). Since they used
fs pulses problems with long pulses are avoided. The high repetition
rate could cause concern, but they checked that lowering it to below 1
MHz did not change the results. Finally, the contribution from the
cascaded nonlinearities should be low: we estimate $|\ns|<10^{-21}~\rm
m^2/W$.

As discussed later in App.~\ref{sec:Wavel-scal-nonl} we use Miller's rule
to convert the nonlinear coefficients to the $\lambda_1=1.03\mic$ that
we use in the simulations in Sec.~\ref{sec:Numer-simul}. This implies
that in the numerics we use $n_{\rm Kerr,11}^I=18.0\times
10^{-20}~\rm m^2/W$, $n_{\rm Kerr,12}^I=6.0\times
10^{-20}~\rm m^2/W$, and $n_{\rm Kerr,22}^I=18.3\times
10^{-20}~\rm m^2/W$. 



\section{Conversion relations}
\label{sec:Conversion-relations}

Often the nonlinear susceptibility is reported in Gaussian cgs units (esu)
instead of the SI mks units. The
conversion between esu and SI is
\begin{eqnarray}\label{eq:chi3-SI-esu}
\chi_{\rm SI}^{(3)}=4\pi\chi_{\rm
  esu}^{(3)}(10^4/c)^2  
\end{eqnarray}
where $c$ is the speed of light in SI
units. The $4\pi$ comes from the Gaussian unit definition of the
electric displacement $\D=\E+4\pi\Pol$, and the $10^4/c$ comes from
converting $\rm statvolt/cm$ to $\rm V/m$.

In most cases the nonlinear Kerr refractive index is used. It is
usually defined as the intensity-dependent change $\Delta n$ in the
refractive index observed by the light
\begin{eqnarray}\label{eq:n2-def}
  n=n_0+\tfrac{1}{2}\Delta n=n_0+n_{\rm
  Kerr}\tfrac{1}{2}|\Ef_0|^2=n_0+n_{\rm Kerr}^I I_0 
\end{eqnarray}
Here $n_0$ represents the linear refractive index, $\Ef_0$ and $I_0$
the input electric field and intensity, respectively. In our case the
total polarization (linear and cubic, in absence of quadratic
nonlinearities) can be written as $P_i=P_i^{(1)}+P_{{\rm
    NL},i}^{(3)}=\varepsilon_0(\varepsilon_i+ \varepsilon_{{\rm
    NL},i})\Ef_i$. Now writing the sum of the linear and nonlinear
relative permittivities as $\varepsilon_i+ \varepsilon_{{\rm
    NL},i}=(n_i+\tfrac{1}{2}\Delta n_i)^2\simeq n_i^2+n_i \Delta n_i$
(here we take $\Delta n_i\ll n_i$) then we can write the change in
refractive index due to the Kerr nonlinearity on the form
\begin{eqnarray}
  \Delta n_i\simeq n_{{\rm Kerr},ii}|\Ef_i|^2+2 n_{{\rm Kerr},ij}|\Ef_j|^2
\end{eqnarray}
When comparing with
Eq.~(\ref{eq:pol-typeI-svea}) we get in SI units \cite{hutchings:1992} 
\begin{eqnarray}\label{eq:n2-SI}
  n_{{\rm Kerr},ij}({\rm SI})=\frac{3}{4n_i}\chi^{(3)}_{{\rm eff},ij}({\rm SI}), \quad i,j=1,2
\end{eqnarray}
Note that the numerical prefactor $3/4$ is the $K$-factor discussed
above. Adopting the intensity notation the change in refractive index
is $ \Delta n_i\simeq 2(n^I_{{\rm Kerr},ii}I_i+2n^I_{{\rm Kerr},ij}I_j)$,
and since in SI units $I_i=\tfrac{1}{2}\varepsilon_0
n_i c|\Ef_{i}|^2$, we get
\begin{eqnarray}
\label{eq:n2I-n2-SI}
  n_{{\rm Kerr},ij}^I({\rm SI})&=&\frac{1}{n_j \varepsilon_0 c}n_{{\rm
    Kerr},ij}({\rm SI})
\\
\label{eq:n2I-chi3-SI}
&=&\frac{3}{4n_in_j \varepsilon_0 c}\chi_{{\rm eff},ij}^{(3)}({\rm SI})
\end{eqnarray}

With Gaussian cgs units we would instead get \cite{hutchings:1992}
\begin{eqnarray}\label{eq:n2-esu}
  n_{{\rm Kerr},ij}({\rm esu})&=&\frac{3\pi}{n_i}\chi_{{\rm
  eff},ij}^{(3)}({\rm esu})\\
\label{eq:n2I-n2-esu}
  n_{{\rm Kerr},ij}^I({\rm esu})&=&\frac{4\pi}{n_j c}n_{{\rm Kerr},ij}({\rm esu})\\
\label{eq:n2I-chi3-esu}
&=&\frac{12\pi^2}{n_in_j c}\chi_{{\rm eff},ij}^{(3)}({\rm esu})
\end{eqnarray}
We have here used that in Gaussian units the intensity is $I_i({\rm
  esu})=(8\pi)^{-1}n_i c|\Ef_i({\rm esu})|^2$. The $K$-factor appears also
in Eq.~(\ref{eq:n2-esu}) as $\tfrac{3}{4}4\pi=3\pi$. Note that $c$ is
still in SI units in these expressions.


The connection between the Gaussian and SI
systems can best be done via Eq.~(\ref{eq:chi3-SI-esu}) and
(\ref{eq:n2I-chi3-SI}) to give \cite{butcher:1990,hutchings:1992}
\begin{eqnarray}\label{eq:n2-esu-SI}
\chi_{{\rm eff},ij}^{(3)}({\rm esu})&=&\frac{n_i n_jc}{120\pi^2}n_{{\rm Kerr},ij}^I({\rm
    SI})\\
\label{eq:n2-I-esu-SI}
  n_{{\rm Kerr},ij}({\rm esu})&=&\frac{n_j c}{40\pi}n_{{\rm Kerr},ij}^I({\rm
    SI})
\end{eqnarray}
where we have used that the SI system defines
$\varepsilon_0c^2=1/\mu_0=10^7/4\pi$ using $c=299\;792\;458~{\rm m/s}$
exactly.

Note that often the definition of the Kerr nonlinear refractive index
is $n=n_0+\Delta n=n_0+n_{\rm Kerr}|\Ef|^2=n_0+n_{\rm Kerr}^I I$ (in Ref.
\cite{bache:2007} we used this notation), which introduces an
additional factor of 2 between $n_{\rm Kerr}$ and $n_{\rm Kerr}^I$,
while the relation between $ n_{\rm Kerr}^I$ and $\chi^{(3)}$ is
unaffected. Thus, working with $\chi^{(3)}$ and $n_{\rm Kerr}^I$ is
the safest because one never has to worry about this factor of 2; as
an example Eq.~(\ref{eq:n2-esu-SI}) is still valid, while with the
alternative definition Eq.~(\ref{eq:n2-I-esu-SI}) becomes $n_{\rm
  Kerr}({\rm esu})=(n_0 c/80\pi)n_{\rm Kerr}^I({\rm SI})$
\cite{butcher:1990}.

\section{Wavelength scaling of the nonlinear susceptibility: Miller's
  delta}
\label{sec:Wavel-scal-nonl}

In the results presented here we account for the wavelength
dependence of the nonlinear coefficients by using
Miller's rule, which states that the following coefficients (the
Miller's delta) are frequency independent \cite{miller:1964}
\begin{eqnarray}\label{eq:Miller-chi2}
  \delta^{(2)}=\frac{\chi_{ijk}^{(2)}}
  {\chi_{ii}^{(1)}\chi_{jj}^{(1)}\chi_{kk}^{(1)}}, \quad  i,j,k=x,y,z
\end{eqnarray}
and we remind that the linear susceptibility is
$1+\chi_{ii}^{(1)}=n_i^2$. A similar relation holds for the cubic
nonlinearity 
\begin{eqnarray}\label{eq:Miller-chi3}
  \delta^{(3)}=\frac{\chi_{ijkl}^{(3)}}
  {\chi_{ii}^{(1)}\chi_{jj}^{(1)}\chi_{kk}^{(1)}\chi_{ll}^{(1)}},
  \quad  i,j,k,l=x,y,z 
\end{eqnarray}
We remark that Miller's delta is based on an anharmonic oscillator
with a single resonant frequency and only gives a ballpark estimate of the
value, and thus is not to be expected to have a large accuracy
(see, e.g., \cite{Bell:1972,Shoji:1997}). However, it has been shown
to work decently for most nonlinear crystals \cite{Alford:2001}.


%

\end{document}